\documentclass[aps,prl,10pt,notitlepage,nofootinbib,tightenlines,floatfix,
twocolumn,superscriptaddress]{revtex4-2}

\usepackage[english]{babel}
\usepackage[margin=50pt]{geometry}

\usepackage[colorlinks=true, allcolors=blue]{hyperref}
\hypersetup{
    colorlinks=true,       
    linkcolor=red,          
    citecolor=magenta,        
    filecolor=magenta,      
    urlcolor=cyan,           
    runcolor=cyan
}
\usepackage{bm,bbm,amssymb,amsmath,amsfonts,amsthm,mathrsfs,MnSymbol,times}
\usepackage{algorithm2e}
\usepackage[normalem]{ulem}

\usepackage{bibunits}
\usepackage{bbm}

\usepackage{graphicx}
\usepackage{amsmath}
\usepackage{physics}
\usepackage{mathtools}
\usepackage{color}
\usepackage{bbold}
\usepackage[utf8]{inputenc}
\usepackage{empheq}
\usepackage{braket}
\usepackage{enumerate}

\newcommand{\eq}[1]{\begin{align}#1\end{align}}
\newcommand{\seq}[1]{\begin{align}\begin{split}#1\end{split}\end{align}}

\newcommand{\bel}{\begin{easylist}[itemize]}
\newcommand{\eel}{\end{easylist}}
\newcommand{\vsigma}{\vec{\sigma}}
\newcommand{\vtau}{\vec{\tau}}
\newcommand{\sx}{\sigma_x}
\newcommand{\sy}{\sigma_y}
\newcommand{\sz}{\sigma_z}
\newcommand{\tx}{\tau_x}
\newcommand{\ty}{\tau_y}

\newcommand{\mrm}{\mathrm}

\newcommand{\mF}{\mathcal{F}}

\renewcommand{\l}{\left}
\renewcommand{\r}{\right}
\DeclareMathOperator*{\argmax}{arg\,max}

\definecolor{ao(english)}{rgb}{0.0, 0.5, 0.0}

\graphicspath{ {figures/} }

\begin{document}

\begin{bibunit}[apsrev4-1]

\title{Suppressing Si Valley Excitation and Valley-Induced Spin Dephasing for Long-Distance Shuttling}

\author{Yasuo Oda}
\affiliation{Department of Physics,
University of Maryland Baltimore County, Baltimore, MD 21250, USA}
\author{Merritt P.~Losert}
\affiliation{Department of Physics, University of Wisconsin-Madison, Madison, Wisconsin 53706, USA}
\author{J.~P.~Kestner}
\affiliation{Department of Physics,
University of Maryland Baltimore County, Baltimore, MD 21250, USA}   
\begin{abstract}
    We present a scalable protocol for suppressing errors during electron spin shuttling in silicon quantum dots. The approach maps the valley Hamiltonian to a Landau-Zener problem to model the nonadiabatic dynamics in regions of small valley splitting. 
    An optimization refines the shuttling velocity profile over a single small segment of the shuttling path, allowing for fast shuttles and ensuring the spin and valley degrees of freedom are unentangled. The protocol reliably returns the valley state to the ground state at the end of the shuttle, after which the spin phase accumulation becomes predictable and is compensated with a single virtual $z$ rotation on the spin.
    The time cost and complexity of the error suppression is minimal and independent of the distance over which the spin is shuttled, and the maximum velocities imposed by valley physics are found to be orders of magnitude larger than shuttling speeds achieved experimentally so far.
    This protocol offers a chip-scale solution for high-fidelity quantum transport in silicon spin-based quantum computing devices.
\end{abstract}

\maketitle

\textit{Introduction.}--
Efficient and coherent spin shuttling has become a promising route for scalable quantum computing in silicon-based systems~\cite{Fujita2017Coherent,Mills2019Shuttling,Yoneda2021Coherent,Jadot2021Distant,Noiri2022shuttling,Boter2022Spiderweb,Zwerver2023Shuttling,Struck2024Spin}. 
Silicon spin qubits benefit from long coherence times due to the weak spin-orbit coupling and the availability of isotopically purified silicon~\cite{Tyryshkin2006Coherence,Burkard2023Semiconductor}. 
However, the natural interaction between spin qubits is the short-range exchange coupling, and experimental devices have inherently limited connectivity.
As a consequence, scalability to the fault-tolerant regime relies on either introducing mediated long-range coupling via large microwave resonators~\cite{Harvey-Collard2022Resonator} or else shuttling spins throughout the device to interact with distant spins~\cite{Seidler2022Conveyor,Langrock2023Blueprint,Xue2024Si,white2024electricalinterconnectssiliconspin}.

Initial considerations of shuttling primarily focused on the charge and spin degrees of freedom. 
Electrons confined in silicon quantum dots, however, present a valley degree of freedom, which arises from the conduction band valley degeneracy in silicon~\cite{Zwanenburg2013Silicon,Friesen2007,Culcer2010,Mi2018}.
It has been widely recognized that the valley degree of freedom inherent to silicon poses a significant challenge for long-range shuttling operations~\cite{Langrock2023Blueprint,Zwerver2023Shuttling}. Valley splitting has been shown to vary significantly even over small distances~\cite{Lima2024Valley}, and efforts to model this fluctuation have considered both deterministic and disorder-dominated regimes~\cite{Tariq2019Effects,Gamble2013Disorder,Ruskov2018Electron,Friesen2007Valley}.
Experimental studies have shown coherent charge shuttling over several microns, but phase coherence has only been demonstrated in closed loops or over shorter distances where valley variability likely plays a less critical role~\cite{DeSmet2024High,Zwerver2023Shuttling,Struck2024Spin}.
In a longer-distance implementation, the spin will inevitably have to be shuttled through regions where the valley splitting is small~\cite{Losert2024Strategies}, leading to transitions between the instantaneous valley eigenstates. This affects the quantum information stored in the spin degree of freedom through the dependence of both the spin exchange coupling and $g$ factor on the valley state~\cite{Hwang2017Impact}. Thus, valley excitations give rise to spin-valley entanglement after some time, which, if left unchecked, induces dephasing of the spin qubits.

Strategies for mitigating valley-induced dephasing during spin shuttling have recently gained attention, either through optimizing interdot tunnel couplings for a bucket-brigade approach \cite{Feng2023Control}, sharpening quantum well interfaces and/or introducing Ge doping to increase valley splitting for a conveyor-mode approach~\cite{Losert2023Practical,Losert2024Strategies}, optimizing the shuttle pulses to enhance adiabaticity~\cite{Losert2024Strategies,Lima2024Superadiabatic} or detouring around locations with dangerously small valley splittings~\cite{Losert2024Strategies}.
Although in many cases these strategies are capable of achieving shuttling fidelities compatible with or higher than single-qubit gate performance, they require advanced capabilities, and one is still left with an infidelity that grows with shuttle distance. 

This paper introduces a simple shuttling protocol designed to suppress valley-related errors over a shuttle of arbitrary distance. 
The main idea is that, as long as the shuttle time is less than the valley relaxation time, the spin-valley dynamics produced by the shuttle are deterministic and do \emph{not} constitute dephasing. Only if the valley state stochastically relaxes while entangled with the spin is the quantum information lost. Thus, one may allow nonadiabatic valley dynamics as long as one then brings the valley back into an eigenstate at the end, preventing spin-valley entanglement, and producing only a deterministic spin rotation. 

\begin{figure}[t]
    \centering
    \includegraphics[width=\linewidth, trim={0cm 0.4cm 13cm 0cm},clip]{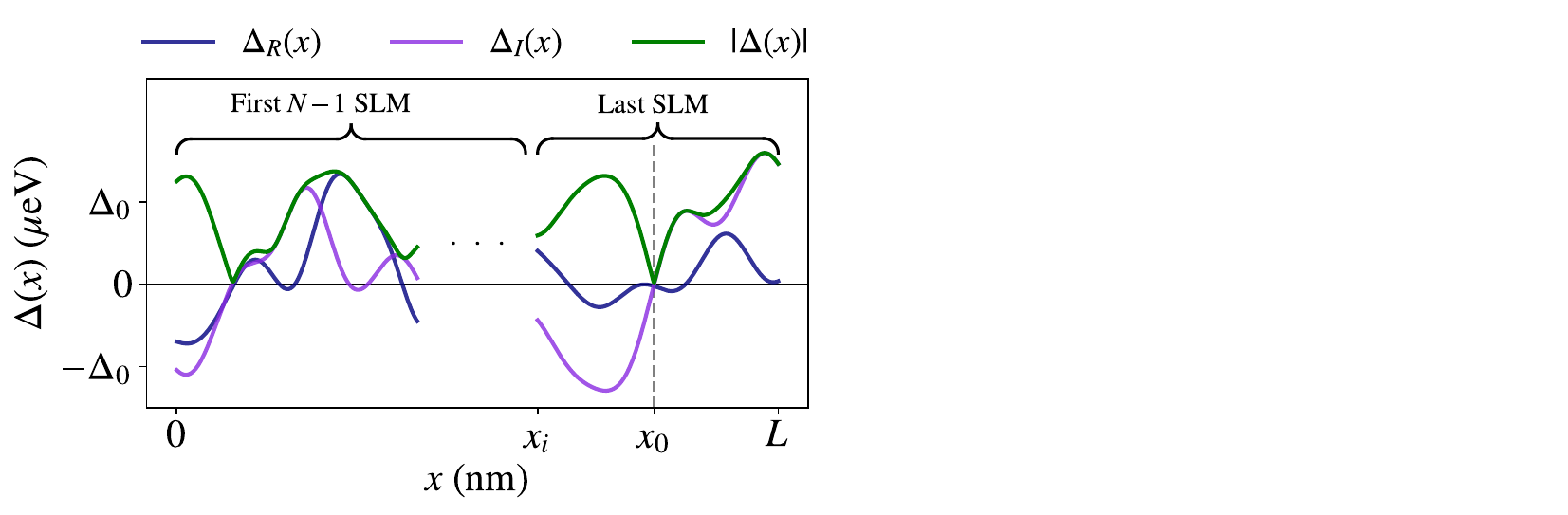}
    \caption{Schematics of real (blue), imaginary (orange), and magnitude (green) of valley landscapes as function of position. We distinguish between the first $N-1$ SLM, through which the electron is shuttled as quickly as possible, and the last SLM, located at $x_0$, over which the velocity profile is optimized. The parameter $\sigma_\Delta$, typically in the order of $100 \,\mu$eV sets the magnitude of the valley splitting fluctuations.}
    \label{fig:main}
\end{figure}

In particular, we split the shuttling into two segments: one segment comprises nearly the entire length of the shuttle path, and the spin is simply moved as fast as possible in this segment, resulting in an uncontrolled but deterministic valley excitation. The other segment is a small region near one of the valley avoided crossings (e.g., the last one in the shuttle path, as in Fig.~\ref{fig:main}), which we refer to as a small local minimum (SLM). Around the selected SLM, we leverage intuition gained from Landau-Zener (LZ) dynamics and apply optimized control strategies to return the valley to the ground state~\cite{Ivakhnenko2023Nonadiabatic,Vitanov1996LandauZener}, 
leaving only a predictable phase accumulation on the spin due to spin-valley coupling. 

This approach has several attractive features. 
Since the optimized part of the shuttle pulse is independent of the total distance, the fidelity can remain high for very long shuttles. 
For instance, a fast shuttle with rate $\sim 500$ m/s (which is still slow enough to avoid orbital state excitations~\cite{Langrock2023Blueprint}), valley relaxation time $\sim 1$ ms, and spin-valley coupling of $\sim 1\,$MHz would allow distances of several mm, opening up chip-scale high-fidelity shuttling.
Even more enticingly in the short term, if one does not have a spatial map of the valley splitting~\cite{Volmer2024Mapping} over the entire shuttle path in order to carry out a theoretical optimization of the ramping pulse \textit{a priori}, our protocol is simple enough and converges quickly enough that one can straightforwardly find the optimal shuttle pulse experimentally \textit{in situ} via a gradient-free Nelder-Mead optimization (or any other optimization method of choice).
Moreover, compared to existing schemes, our approach presents both software and hardware advantages: we trade the cost of valley mapping for optimization complexity, but does not introduce any additional hardware modifications.

Regardless of the optimization specifics, valley-induced dephasing during shuttling can be efficiently removed by changing the control parameters, and as such it constitutes a calibration issue rather than a source of decoherence.
Thus, this strategy promises to enable chip-scale electron shuttling, particularly suited for conveyor-mode settings where the shuttling speeds can be tuned in accordance to the current proposal.

\textit{Spin-shuttling model}.--
Following Ref.~\cite{Losert2024Strategies}, we consider the spin-shuttling dynamics in the valley-spin coupling model described by the minimal Hamiltonian
\seq{
\label{eq:H_SV}
H(x) = \frac{\omega_B}{2} \sz + \frac{1}{2} \vec{\Delta}(x) \cdot \vtau \otimes \l( 1 +  \frac{\delta \omega_B}{|\Delta (x)|} \sz \r).
}
Here, $\omega_B$ is the Zeeman splitting energy of the spin degree of freedom, $\vec{\Delta}=(\Delta_R,\Delta_I)$ where $\Delta_R(x)$ and $\Delta_I(x)$ are the real and imaginary parts of the complex-valued intervalley coupling $\Delta(x)=\Delta_R(x)+i\Delta_I(x)$, and $\delta \omega_B$ is the difference in Zeeman splittings for the two valley states.
The Hamiltonian includes both spin and valley degrees of freedom, where $\vtau=(\tx,\ty)$ and $\vsigma=(\sx,\sy,\sz)$, are Pauli operators acting on the valley and spin, respectively. 
The variations in intervalley splitting lead to regions where transitions between valley states can occur.
Owing to the difference in Zeeman splittings for each valley state, this leads to spin-valley entanglement in the shuttling process, in turn causing dephasing and degraded fidelity upon stochastic collapse of the valley state.

We begin by temporarily neglecting the backaction of the spin-valley coupling on the valley evolution, thus focusing on the dynamics of the valley states alone. 
The spin contribution is weak compared to typical magnitudes of intervalley couplings, and as shown in the Supplemental Material \cite{SM} exerts minimal influence on valley state evolution. 
Thus, the pure valley Hamiltonian simplifies to
\begin{equation}
\label{eq:H_V}
    H_\mrm{V}(x) = \frac{\Delta_R(x)}{2} \tx + \frac{\Delta_I(x)}{2} \ty,
\end{equation}
where the splitting between the valley states is $E_V=|\Delta|$.
The main degree of freedom is the shuttling velocity $v(t)$, which enters the Hamiltonian via the standard relationship $x(t)=\int_{0}^t v(s) ds$, and naturally gives rise to the time-dependent valley Hamiltonian $H_\mrm{V}(t)=H_\mrm{V}(x(t))$ and the time-ordered propagator $U_\mrm{V}(t_2;t_1)=\mathcal{T}_+ e^{-i\int_{t_1}^{t_2} H_\mrm{V}(s) ds}$. 
This approach allows us to focus on constructing a velocity profile for the electron that minimizes valley transitions while ensuring that the electron is shuttled efficiently. 

\textit{Overview}.--
The primary goal of the proposed shuttling protocol is to nonadiabatically transport the electron from position $x=0$ to $x=L$ across multiple SLMs such that it ends up in the ground valley eigenstate. 
As shown in the Supplemental Material~\cite{SM}, a sufficiently fast shuttle leads to negligible spin-valley entanglement during the shuttle and acts only on the valley state.

The protocol starts by identifying the location of one SLM. For the sake of simplicity we will take this to be the last SLM on the path, located at a position $x=x_0$, although any one would work. Then, the electron is shuttled at the maximum possible speed from $x=0$ to a point $x_i$, chosen such that $|\Delta(x_i)| \gg |\Delta(x_0)|$, as shown in Fig.~\ref{fig:main}.
The nonadiabatic traversal of the SLMs along the way transforms the valley state into $\ket{\psi(x_i)}=\cos(\theta_i/2) \ket{0} + e^{i\phi_i} \sin(\theta_i/2) \ket{1}$ with Bloch angles $(\theta_i,\phi_i)$, which generally differs from an instantaneous valley eigenstate.

This allows us to treat the electron’s passage through regions of small valley splitting as a controlled LZ transition by tuning the velocity profile.
Denoting $t_i$ such that $x(t_i)=x_i$, the goal of the optimization is to find the velocity profile $v(t)$ for $t_i\leq t\leq t_f$ with the condition that $\ket{\psi(L)}=\ket{\psi_-}$, by which we denote the instantaneous valley ground state satisfying $H_\mrm{V}(L)\ket{\psi_-}=-\tfrac{1}{2}|\Delta(L)|\ket{\psi_-}$.
More concretely, we aim to maximize the state fidelity 
\seq{
\label{eq:F_V}
\mathcal{F}_\mrm{V}(t) &= |\bra{\psi_-} U_\mrm{V}(t;t_i) \ket{\psi(x_i)}|^2,
}
at $t=t_f$, so the optimization problem is defined by
\seq{
\label{eq:opt_prob_Fv}
v_\mrm{opt}(t) &= \argmax_{v(t)} \mathcal{F}_\mrm{V}(t_f),
}
which yields an optimized velocity profile that helps suppress valley-induced dephasing by driving the valley to its ground state.
 
\textit{Velocity profile ansatz}.--
Arguably, the simplest control scheme consists of a constant optimal velocity shuttle for the last segment of the shuttle. 
However, as shown in the Supplemental Material~\cite{SM}, while this can be partially effective, it is too constrained to reliably produce high fidelity shuttles.
In this work, we draw inspiration from the LZ problem, where the physics become markedly distinct as one approaches the anticrossing point, where state transitions can occur. 
As the next most simple strategy, one could select different control parameters for the regions near and far from the SLM to achieve the desired evolution.
Motivated by these ideas, we choose to segment the shuttling process into three regions of constant velocity, namely
\seq{\label{eq:v}
v(t) = \begin{cases}
    v_1 & \mathrm{if}\quad |t-t_0| > \tau,\\
    v_2 & \mathrm{if}\quad |t-t_0| < \tau,
\end{cases}
}
where we imposed a symmetric condition on the velocities for simplicity.
Intuitively, $v_1$ is chosen as large as possible to minimize the duration of the shuttle, whereas $v_2$ is constrained by requiring that the electron spends enough time near the SLM to induce the necessary valley state rotation. Note that this ``fast-slow-fast'' strategy based on the piecewise-evolution approximation (PEA) is in some sense opposite to the ``slow-fast-slow'' approach discussed in the context of the adiabatic-impulse approximation~\cite{Kayanuma1997}.

Here, the state is allowed to change as opposed to remaining in the ground state as in the adiabatic evolution, while any changes in the valley are undone to return it to the ground state at the end of the shuttle.
The dynamics induced by Eq.~\eqref{eq:v} are simple enough that the state evolution can be calculated explicitly near the SLM where the valley physics resembles a LZ problem.
We denote this framework by PEA, and it is developed in detail in the Supplemental Material~\cite{SM}.
In this regime, the optimal velocities $(v_1,v_2)$ necessary to return the valley to the ground state can be found analytically, and can be used as initial seed for an optimization that would account for any model inaccuracies.

\textit{Optimization protocol}.--
If one knows the valley landscape, and hence the initial state parameters $(\theta_i,\phi_i)$, the velocity profile obtained from the PEA is a good initial guess for a numerical gradient descent optimization of the actual valley state fidelity Eq.~\eqref{eq:F_V} at final time $t_f$. 
In practice, however, obtaining precise knowledge of the valley landscape may be costly. Our protocol then does not assume any knowledge of the valley landscape, and we use the piecewise constant velocity ansatz of Eq.~\eqref{eq:v} for an experimental gradient-free optimization of $(v_1,v_2,t_0,\tau)$.
Intuitively, by tuning $t_0,\tau$, the optimizer is given enough flexibility to find the location of the SLMs.
As discussed in detail in the Supplemental Material~\cite{SM}, convergence is aided by choosing an initial seed informed by the statistics of optimized pulses over an ensemble of initial angles and random valley landscapes with properties that statistically reflect the physics of the device. 
This pretraining step can be performed offline or on a database of previously optimized chips, and thus does not incur additional computational cost.

Experimentally, one applies an arbitrarily fast shuttle until reaching the selected optimization region, at which point one changes the ramping speed to the initial seed.
The valley ground state probability is then measured at the endpoint and the shuttle pulse updated via the Nelder-Mead algorithm~\cite{Nelder1965}.
This whole process is to be iterated until the desired ground state fidelity is reached (otherwise the process can be restarted with a different random three-piece seed).
Finally, the deterministic spin rotation along $z$ induced by the coupling to the valley after the shuttle would be measured and compensated virtually via a local oscillator phase update~\cite{McKay2017_virtual}.

\begin{figure}[t]
    \centering
    \includegraphics[width=0.8\linewidth, trim={0.7cm .4cm 0cm 0cm},clip]{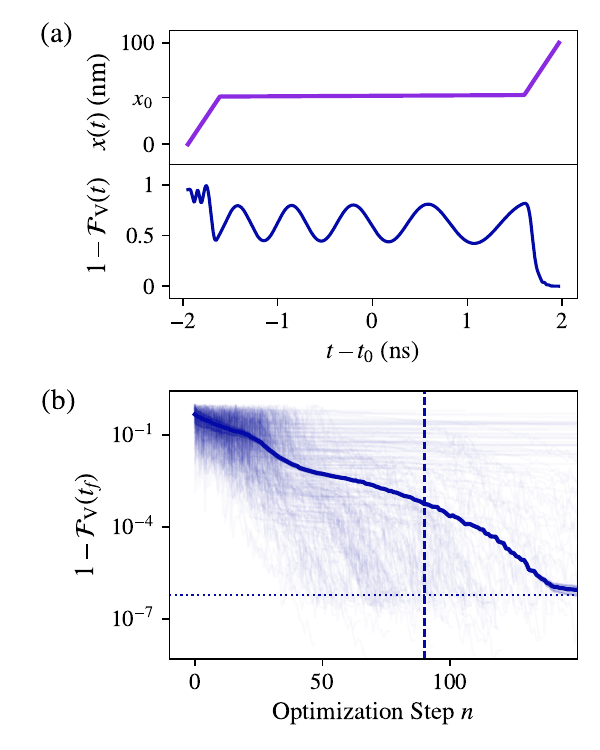}
    \caption{(a) Example of optimized electron shuttling (top) and infidelities (bottom) as functions of time, obtained from optimizing on the last segment of the landscape shown in Fig.~\ref{fig:main}, for an initial state with angles $(\phi_i, \theta_i) = (\pi/4, \pi/2)$, over a distance of $100$\,nm at an average speed of $\sim$35\,m/s. 
    (b) Median state infidelity [Eq.~\eqref{eq:F_V}] at final shuttling time (solid lines) as function of optimization step up to 150 iterations.
    First interquartile ranges are shown as error bars in shaded region.
    In addition, 500 individual optimizations are shown in the background.
    Vertical dashed line represents median values of total number of iteration steps. Horizontal dotted line marks achieved median infidelities at the end of the shuttle.
    }
    \label{fig:results}
\end{figure}

\textit{Results}.--
A representative example of the valley fidelity obtained from optimization can be found in Fig.~\ref{fig:results}(a) for the valley landscape near the last SLM shown in Fig.~\ref{fig:main}.
Here, it can be seen that the dot spends most of the shuttling around the SLM as the electron is shuttled quickly to that point at a speed $300\,$m/s, while around the SLM dropping to within the $\pm 1\,$m/s range, reflected by the sharp change in slope in the top panel of Fig.~\ref{fig:results}(a).

We tested the optimization protocols on 100 different randomly generated landscapes, generated following Ref.~\cite{Losert2024Strategies}, and 100 initial states given by $(\phi_i,\theta_i)$, for a total of $10^4$ test cases.
For these optimizations, we consider symmetric dots of size $a=14\,$nm. 
The valley coupling is determined by the local atomic disorder in the device. Since the dot averages over the disorder, the dot size sets the length scale of valley coupling fluctuations.
After determining the dot size, we chose a shuttling distance of $100\,$nm in order for at least one SLM to be present, although sometimes more than one SLM is encountered. 
The maximum velocity was constrained to $< 10^3\,$m/s to avoid orbital excitations (see Ref.~\cite{Langrock2023Blueprint}).
The time evolution given by the operator $U_\mrm{V}$ was implemented in Qutip~\cite{Johansson2013_qutip}.
More details regarding the generation of landscapes and choice of optimization parameters can be found in the Supplemental Material~\cite{SM}.

Figure~\ref{fig:results}(b) shows the infidelity vs number of optimization iterations for both methods.
Over the set of $10^4$ different cases, we find a median valley state infidelity $1-\mathcal{F}_\mrm{V}(t_f)$ of $4\times 10^{-7}$, demonstrating that one can reliably find solutions to arbitrary accuracy without any knowledge of the valley landscape.
The median number of iterations it took the optimizer to find solutions with a tolerance of $1-\mF_\mrm{V}(t_f)<10^{-6}$ was 90.

When detailed information of the valley landscape is accessible, it is possible to differentiate the Hamiltonian explicitly, and thus execute a gradient-based optimization instead. 
As shown in the Supplemental Material~\cite{SM}, this ``valley-informed'' approach yields some benefits compared to the ``valley-agnostic'' method presented in this section, particularly in a reduction in the number of optimization steps needed as the location of the SLM is known, and only the velocities must be optimized.

\textit{Discussion}.--
We now address several points regarding the practical implementation, as well as sources of error we have not included in the optimization.

Although in this work we chose to optimize passage through the last SLM on the shuttle path for simplicity, an analogous approach can be easily formulated for any intermediate SLM instead.
Utilizing this approach introduces the additional challenge of not knowing with certainty if the optimization is being performed over just one or even no SLM. 
This challenge can be readily addressed, however, by including the shuttled distance as a new optimization variable, without increasing the complexity of the optimization problem substantially.
The solution is then equivalent to the PEA derived in this work, but targeting instead of the ground state a state that compensates for the nonadiabatic valley evolution up to that point and precompensates for the remainder following that point.
This strategy can also be used when a SLM is present but unusable, e.g.,~due to the presence of a spin-valley hot spot in its vicinity, where shuttling slowly near the hot spot will induce fast decoherence~\cite{Yang2013}. In this case, the optimizer would initially fail to find a solution, then choose a different region instead, shuttling quickly past the previously problematic hot spot and minimizing its deleterious effect.

Charge noise may change the shuttle path and consequently the valley landscape seen by the electron. 
In particular, $1/f$-type noise has been observed at kHz--MHz frequencies~\cite{Connors2022,ye2024characterizationindividualchargefluctuators,Yoneda2023,Jock2022}, but with a much lower amplitude that only perturbs the dot position by $\lesssim \delta x \sim 10\,$pm~\cite{Yoneda2017}.
As mentioned earlier, the size of the dot sets typical correlation lengths for effective valley landscapes, 
which means a typical perturbation will be of the order of $\delta \Delta \sim \sigma_\Delta \times (\delta x/a) \approx 0.1\,\mu$eV. For the fast shuttles introduced in this work of duration $T_\mrm{opt} \sim 4\,$ns, we obtain an estimated loss of fidelity of $\delta\mathcal{F}_V \sim (\delta \Delta \times T_\mrm{opt}/2\pi)^2 \approx 2\times 10^{-4}$, which is in the order of currently achievable single-qubit gate fidelities, and consequently we do not expect charge noise to become a limiting factor.
We refer to the Supplemental Material~\cite{SM} for supporting numerical analysis. Similarly, we expect that effects in valley splitting due to changes in the electron wave function radius and overlap with the SiGe barrier will have a comparable effect.

It is worth noting that spin-orbit and valley-orbit couplings may induce hybridization of these degrees of freedom, leading to potential new decay channels of spin and valley through the fast dephasing of orbital states. Although these effects are maximized in the presence of hot spots, regions of valley splitting large enough to induce valley-orbit hot spots are rare. In addition, the large separation of energy scales between the valley and orbital excitations ($\sim$meV) and valley-orbit coupling ($\sim\mu$eV), combined with the fast shuttles proposed here, minimize such hybridization. Although orbital dephasing times have been estimated to be as fast as $60$~ps in some cases~\cite{MacQuarrie2020}, for an orbital dephasing time of 1~ns~\cite{Gorman2005,Mi2017} we estimate that the valley coherence time would remain in the millisecond range, and as a consequence we neglect it in this work.

Although localized charge traps and atomic steps may induce discontinuities in the valley landscape (notably in its phase), these discontinuities are expected to be static and thus will be accounted for in the optimization. In addition, the effect of signal uncertainties, particularly timing uncertainties on the optimized shuttles are shown in the Supplemental Material~\cite{SM} to be minor.

Lastly, we show in the Appendix that the optimization is robust to measurement errors and finite shot noise. In addition, note that with $\sim$ 100 shots at a single-shot measurement time under $1\,$ms~\cite{Elzerman2004} and $\sim$ 100 iteration steps, we expect the optimization process to take $\sim 10\,$s.
Since near-optimal results can be quickly achieved even with noisy measurements and a gradient-free optimization method, this analysis provides optimism regarding the applicability of this method on hardware.

\textit{Conclusions}.--
The protocol we have proposed in this work achieves efficient suppression of valley-induced spin dephasing by leveraging intuition obtained from the LZ transition. 
We perform an optimization process to minimize valley transitions: first by mapping the linearized valley Hamiltonian to a LZ problem and solving it using the PEA, and then, by refining the velocity profile through gradient-free optimization in the original valley frame.

Thus, by adjusting the shuttling velocities in a simple way over a small segment of the path, a rapid, long-distance shuttle over regions of small valley splitting can achieve suppression of valley-induced dephasing independent of the shuttle distance. 
We highlight that, although the optimization introduced in this work is remarkably effective and simple, we regard as the main result of this work the notion that valley-induced dephasing during shuttling does not constitute decoherence, but rather can be efficiently calibrated out.
This protocol paves the way for chip-scale coherent transport in silicon-based quantum processors.
Future work may explore further extensions of the model by incorporating other sources of dephasing, such as deterministic spin-orbit coupling and charge noises, or adapting for shuttling in Ge quantum dots, where the $g$ tensor changes in space~\cite{vanRiggelenDoelman2024}, rather than the valleys.

\acknowledgments

The authors acknowledge support from the Army Research Office (ARO) under Grant No. W911NF-23-1-0115, and thank John H. Caporaletti for useful discussions. 

While finalizing this manuscript we became aware of a protocol introduced in Ref.~\cite{David2024Long} which is somewhat similar to our approach. However, we remark upon key differences: it uses automatic differentiation rather than a gradient-free algorithm; it does not use a LZ-inspired ansatz; it optimizes over the full shuttle distance; and, notably, it does not return the valley to its ground state.

\putbib[references]

\end{bibunit}

\clearpage                 

\onecolumngrid             
\rule{\textwidth}{1.3pt}
\begin{center}
  \textbf{\large End Matter}   
\end{center}
\vspace{0.5cm}
\twocolumngrid             

\noindent\textit{Appendix: Optimizing in the presence of shot and measurement noise}---When the optimization is experimentally performed on hardware, the measured valley state estimated quantities will inevitably suffer from different sources of noise.
To test the efficacy of the optimization when measurement noise is included, we incorporate stochastic contributions to the objective function.
We model shot noise by randomly sampling state projections $q_i\in\{0,1\}$ with probabilities given by $\{ \mF_\mrm{V}(t_f),1- \mF_\mrm{V}(t_f)\}$, respectively.
Then, we define the finite sampling fidelity $\overline{f}_M=1-\frac{1}{M}\sum_{i=1}^M q_i$, where $M$ is the number of shots, which approximates the noiseless fidelity $\mF_\mrm{V}(t_f)=\overline{f}_M+O(1/\sqrt{M})$.
The influence of shot noise will be more significant for superposition states, whereas for computational basis states it will be insignificant. 
Thus, shot noise is expected to only affect noticeably the initial part of the optimization.

We also allow for measurement noise by assuming that the wrong state $q_i$ was detected with uniform probability $p_\mrm{err}$, so that the noisy measured fidelity is
\seq{
\label{eq:noisy_fid}
F_{m}\l(\overline{f}_M\r) =\overline{f}_M \times \l(1 - \frac{m}{M}\r) + \l(1-\overline{f}_M\r) \times \frac{m}{M},
}
where $m$ is a random integer drawn from a Poisson distribution with mean $p_\mrm{err} \times M$.
As opposed to shot noise, measurement noise will contribute significantly only at the end of the optimization, where the average fidelity becomes $\langle F_{m}(1)\rangle=1-p_\mrm{err}$, while for equal superposition states Eq.~\eqref{eq:noisy_fid} has a fixed point, namely, $\langle F_{m}(1/2)\rangle=1/2$.
A different random $m$ is drawn for each call of $F_m$.

\begin{figure}[t]
    \centering
    \includegraphics[width=\linewidth, trim={0cm .45cm 0cm 0cm},clip]{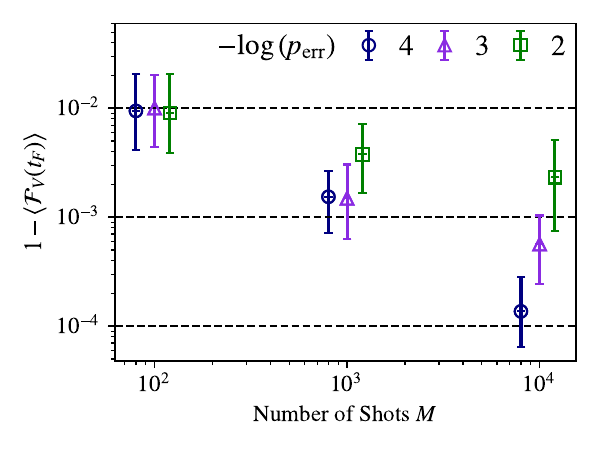}
    \caption{Median state infidelities [Eq.~\eqref{eq:F_V}  of the main text] obtained from optimizing the noisy fidelity [Eq.~\eqref{eq:noisy_fid}] as described in the text, averaged over 600 different combinations of landscapes and angles.
    Error bars are given by first interquartile ranges.
    Measurement error per shot of $p_\mrm{err}\in\{10^{-4},10^{-3},10^{-2}\}$.
    Horizontal dashed lines represent the boundary set by measurement noise, where the infidelity is equal to $p_\mrm{err}$. 
    The number of shots $M\in\{10^2,10^3,10^4\}$ have been shifted slightly for clarity.
    }
    \label{fig:appendix}
\end{figure}

Optimizing with the noisy measured fidelity Eq.~\eqref{eq:noisy_fid} as the objective function using Nelder-Mead we obtain the state infidelities shown in Fig.~\ref{fig:appendix}. 
Each combination of $(p_\mrm{error},M)$ is optimized over a subset of 300 landscapes and initial angles, randomly sampled from the $10^4$ ones used previously in Fig.~\ref{fig:results}(b) of the main text.
The optimizer is able to approach the ground state when using more than 100 shots, reaching the target set by measurement error $1-\braket{F_m} \sim p_\mrm{err}$.
Using the noisy objective function did not require more iterations; the average number of iterations was 69 (less than the noiseless case, but only because the infidelity saturates at a higher level).

\newpage

\appendix

\begin{bibunit}[apsrev4-1]

\newpage
\onecolumngrid

\vspace{1cm}
\newpage

\begin{center}
\textbf{\large Supplemental Material}
\end{center}

\section{Spin-Valley Entanglement on Fast and Long Shuttles}

In this section, we consider an electron moving at constant velocity $v$ along $x(t)=vt$, where we focus on the dynamics that couple spin and valley, given by
\seq{
H_{SV}(x) = \frac{\delta\omega_B}{2}\; \sigma_z\,\frac{\Delta_R(x)\,\tau_x + \Delta_I(x)\,\tau_y}{\abs{\Delta(x)}},
\qquad \Delta(x)=\Delta_R(x)+i \Delta_I(x),
}
where $\qquad \abs{\Delta(x)}=\sqrt{\Delta_R(x)^2+\Delta_I(x)^2}$ and $\phi(t) \equiv \arg\big[\Delta(x(t))\big]$ are the valley amplitude and phase. With the valley phase, we write the explicitly time-dependent Hamiltonian $H_{SV}(x(t)) = \frac{\delta\omega_B}{2}\; \sigma_z \left( \cos\phi(t)\,\tau_x + \sin\phi(t)\,\tau_y \right)$.

The time-evolution induced by $H_\mrm{SV}$ is given by the time-ordered propagator $U(T)=\mathcal{T}_+ e^{-i\int_0^T H_\mrm{SV}(x(t))dt}$. This propagator can be studied perturbatively using a Magnus expansion~\cite{Magnus1954}, where we write $U(T)=\exp[\sum_{n=0}^\infty\Omega(T)]$. Taking the second order approximation, we compute the first two terms of the Magnus series
\seq{
\Omega_1(T) &= -i \int_0^T \! H_{SV}(x(t))\,dt \\
&= -i\,\frac{\delta\omega_B}{2}\; \sigma_z \int_0^T \big(\cos\phi(t) \,\tau_x + \sin\phi(t)\,\tau_y \big) dt,\\
\Omega_2(T) &= -\frac12 \int_0^T\! dt_1 \int_0^{t_1}\! dt_2\; [H_{SV}(x(t_1)),H_{SV}(x(t_2))] \\
&= -i\,\frac{\delta\omega_B^2}{4}\; \left[\int_0^T\! dt_1\!\int_0^{t_1}\! dt_2\; \sin\!\big(\phi(t_2)-\phi(t_1)\big)\right] \tau_z,
}
where in $\Omega_2(T)$ we used $\sigma_z^2=\mathbbm{1}$ and $[\sigma_z,\tau_j]=0$.
Note that $\Omega_1(t)$ depends on integrals of $\cos\phi(t),\sin\phi(t)$, which oscillate quickly and randomly, and, as we argue below, will average out to a large extent.
In addition, $\Omega_2(T)$ does not contain a spin operator, and thus if $\Omega_1(T)\approx0$ then $\Omega_2(T)$ will not contribute to the spin evolution, and is consequently ignored in the remaining of this discussion. 

Focusing on $\Omega_1(T)$, we define $\Phi(T)=\int_0^T \big(\cos\phi(t) \,\tau_x + \sin\phi(t)\,\tau_y \big) dt = ||\Phi(T) || \tau_n$, with
\seq{
||\Phi(T)|| &= \sqrt{\l(\int_0^T \cos\phi(t) dt\r)^2+\l(\int_0^T \sin\phi(t) dt\r)^2}, \\
\tau_n &= \frac{1}{||\Phi(T)||}\l(\l[\int_0^T \cos\phi(t) dt\r] \tau_x +\l[\int_0^T \sin\phi(t) dt] \r]\tau_y\r),
}
satisfying $||\tau_n||=1$.
Then, we write the evolution operator induced by $\Omega_1(T)$ as,
\seq{
U_1(T) &= e^{\Omega_1(T)} = \exp\l(- \frac{i}{2} \delta\omega_B \sigma_z \Phi(T)\r) \\
&= \cos\l(\delta\omega_B||\Phi(T)||\r) I -i \sin\l(\delta\omega_B||\Phi(T)||\r) \sz \tau_n,
}
which approximates $U(T)\approx U_1(T)$ to leading order in $\delta\omega_B T\ll1$.
Using the trace norm, we quantify the distance between this propagator and the identity via the operator fidelity,
\seq{
F(T) &= \frac{1}{4}\abs{\tr U_1(T)}^2 = \cos\l[\delta\omega_B||\Phi(T)||\r]^2 \\
&\approx 1 - \frac{(\delta\omega_B T)^2}{2} I(T),
}
where a second order approximation was taken, and we defined $I(T)=||\Phi(T)||^2/T^2$.
Note that, if $\Delta(x)$ is constant in space, we obtain $I(T)=1$, which is also its maximum value where the spin-valley phase accumulates coherently. Thus, since $\Delta_{R,I}(x)$ will in general take random positive or negative values, the infidelity $1-F(T)$ is suppressed by a factor of $I(T)$, compared to the naive expression of a coherent phase accumulation due to a constant $\Delta(x)$.

More explicitly, we can rewrite the integrals in $I(T)$ in terms of $x=v t$, which is particularly useful when trying to understand the effect of velocity in traversing a fixed distance $L=v T$. Using $dt = dx/v$, 
\seq{
I(T) &= \frac{1}{L^2}\l[ \l(\int_0^L \cos\psi(x) dx\r)^2+\l(\int_0^L \sin\psi(x) dx\r)^2 \r] \\
&= \frac{1}{L^2} \int_0^L \int_0^L \cos\l[\psi(x)-\psi(x') \r] dx dx',
}
where in addition we defined $\psi(x)\equiv \arg \Delta(x) = \phi(t/v)$.
These expressions can be computed explicitly using $\mrm{Cov}[\Delta_R(x),\Delta_R(x')]=\mrm{Cov}[\Delta_I(x),\Delta_I(x')] = \sigma^2 e^{-(x-x')^2/2 a^2}/2$~\cite{Losert2024Strategies}, where $a$ is the size of the dot (assumed circular for simplicity) and $\sigma^2$ is the variance of the Gaussian random variables $\Delta_R,\Delta_I$.
Since $\Delta(x),\Delta^*(x)$ are a jointly Gaussian circular pair satisfying $\mathbb{E}[\Delta(x)\Delta^*(x')]=\sigma^2 e^{-(x-x')^2/2 a^2}$, we have from random wave theory~\cite{Goodman2020} that the phase difference has a wrapped-Cauchy distribution, from where we can derive $\mathbb{E}\l[e^{i(\psi(x)-\psi(x'))}\r]=e^{-(x-x')^2/2 a^2}$.
We can use this to compute the expectation value 
\seq{
\mathbb{E}\l[I(T) \r] &= \frac{1}{L^2}\int_0^L\int_0^L e^{-(x-x')^2/2a^2} dx dx' \\
&\overset{a\ll L}{\approx} \frac{\sqrt{2\pi}a}{L}.
}
Lastly, we estimate the average infidelity 
\seq{
\label{eq:long_shuttle_fid}
1-\mathbb{E}\l[F(T)\r] &\approx \frac{(\delta\omega_B L)^2}{2 v^2} \frac{\sqrt{2\pi}a}{L} \\
&= \sqrt{\frac{\pi}{2}} \frac{\delta\omega_B^2 L a}{v^2},
}
which grows only linearly in traversed distance $L$. 

\begin{figure}[h]
    \centering
    \includegraphics[width=0.5\linewidth, trim={0cm 0cm 0cm 0cm},clip]{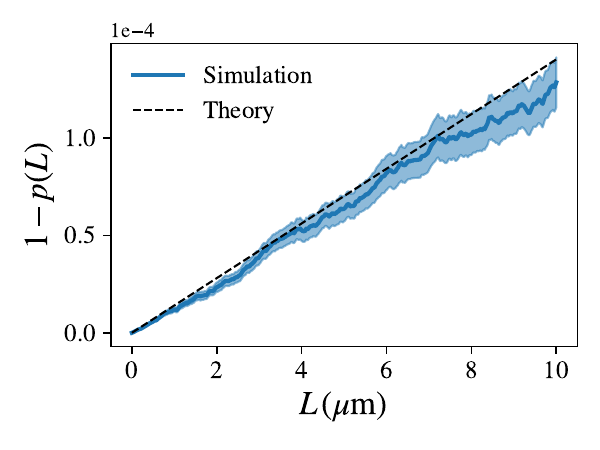}
    \caption{Comparison between simulated and theoretically estimated spin state fidelity for a $L=10\,\mu$m shuttle at $v=500\,$m/s and coupling strength $\delta\omega_B=10\,$MHz. The simulated state survival probability is averaged over 100 different valley landscape realizations (blue), with error bars shown as shaded blue region representing standard error of the mean, but are barely visible. The theoretical prediction from Eq.~\eqref{eq:long_shuttle_fid} is shown as dashed line, showing excellent agreement with the simulation average. 
    }
    \label{fig:long_shuttle}
\end{figure}

We compare this prediction with numerical simulations of fast shuttles in Fig.~\ref{fig:long_shuttle}. Here, the spin state was initialized on the $\ket{+}=(\ket{0}+\ket{1})/\sqrt{2}$ state, while the valley was initialized in its ground state. The spin-valley state survival probability $p(x)=|\bra{+,0}U(x/v)\ket{+,0}|^2$ was computed as a function of time for a $10\,\mu$m shuttle (shown in blue), obtaining excellent agreement between simulation and theory.
In particular, this suggests that for an infidelity of $10^{-3}$, dot size $a=14\,$nm, shuttle of velocity $500\,$m/s, and a coupling strength of $\delta\omega_B=1\,$MHz, the electron could be shuttled for up to $L=10\,$mm, opening up chip-scale high-fidelity shuttling.

\subsection{Spin-Valley Entanglement During Optimized Segment}

\label{app:spin_backaction}

\begin{figure}[h]
    \centering
    \includegraphics[width=.6\linewidth]{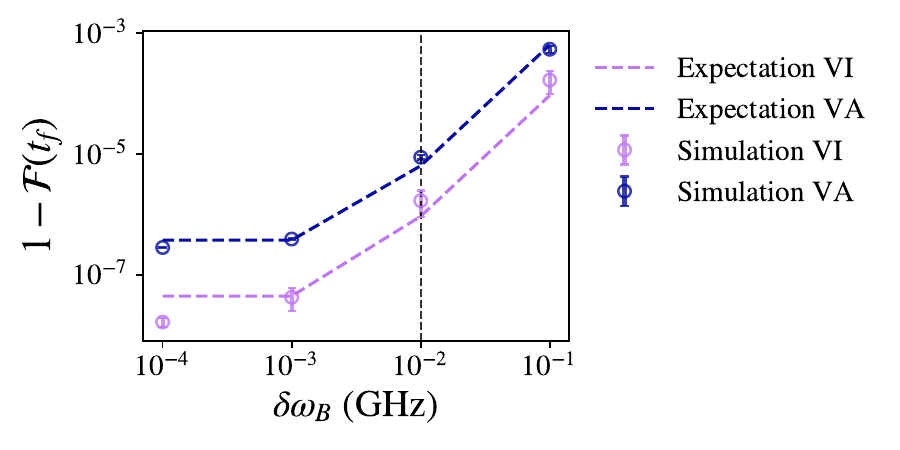}
    \caption{Spin backaction on valley evolution, as measured by median state infidelity Eq.~\eqref{eq:2Qinf} (empty circles) over a subset of 1000 different testcases, drawn randomly from the dataset considered in the main text. 
    Error bars represent first interquartile ranges.
    The dashed vertical line marks the typical value of $\delta\omega_B\sim 10^{-2}\,$GHz, below which the infidelities saturate to the uncoupled infidelities obtained from optimization.
    Dashed line represents the expectations of infidelity contribution from the spin-valley coupling, given by the maximum between $(\delta\omega_B T_\mrm{opt}/(2\pi))^2/2$ and the uncoupled infidelities.}
    \label{fig:spin_backaction}
\end{figure}

Here, we analyze numerically the effect of the spin degree of freedom on the valley dynamics.
We numerically simulate the spin-valley Hamiltonian Eq.~\eqref{eq:H_SV}, by choosing the velocity profile optimized shown in Fig.~\ref{fig:results} of the main text.
We calculate the ground state probability as a function of time after tracing out the spin,
\seq{\label{eq:2Qinf}
\mathcal{F}(t)=\bra{0}\tr_{S}\l( U(t) \rho_i U(t)^\dagger \r)\ket{0},
}
where $\rho_i=\ket{\psi(x_i)}\bra{\psi(x_i)}$ is the initial density matrix state. The full propagator is computed by $U(t)=\mathcal{T}_+ e^{-i\int_{0}^{t} H(s) ds}$, where $H(t)=H(x(t))$.
We use the velocity profiles obtained from the VI and VA optimizations described in the main text to compute $x(t)$, used to compute the dynamics.
The results of these computations are shown in Fig.~\ref{fig:spin_backaction}.

We vary $\delta \omega_B$ over four orders of magnitude, to analyze the dependence of the backaction on the coupling strength.
For each case, we initialize the spin state on a random state of the Bloch sphere.
The results align well with order of magnitude expectations, shown in solid lines.
We highlight a typical value of $\delta\omega_B\sim 10^{-2}\,$GHz~\cite{Losert2024Strategies}, where the mean infidelity is below $10^{-4}$.
In addition, in the weak coupling regime, we expect the influence of the spin on the valley to be proportional to $(\delta \omega_B T_\mrm{opt})^2$, where $T_\mrm{opt}=t_f-t_i$ is the duration of the optimized segment of the shuttle.
In Fig.~\ref{fig:spin_backaction} we show this for reference, for the mean shuttling times of $1.7$ and $4.5\,$ns for VI and VA optimizations, respectively.
We compare the results of simulation with the expectation that the infidelity will be dominated by the spin-valley coupling when the pure valley infidelity is lower than $(\delta \omega_B T_\mrm{opt})^2$, and vice versa.
This explains the saturation observed for $\delta\omega_B<10^{-2}$~GHz, saturating to the values of $4.5\times 10^{-8}, 3.8\times 10^{-7}$ corresponding to the valley-only infidelities obtained from optimization. We observe excellent agreement between simulation and this expectation.

\section{Constant Speed Shuttle Optimization}

In this section, we explore the strategy of shuttling at constant speed~~ \cite{Langrock2023Blueprint,Losert2024Strategies,Lima2024Superadiabatic}.
In the low-speed regime, the adiabatic passage allows the valley to remain in its instantaneous ground state throughout the duration of the shuttle~\cite{Lu2005,Li2023}.
Although this could be done in practice, it would likely take too long compared to estimated valley lifetimes~\cite{Yang2013}.
To try to get past this limitation, we study the problem of shuttling as quickly as possible until the vicinity of the electron destination (here taken to be the last 100~nm of the shuttle), and optimizing the electron speed over that last region.
The goal of the optimization is to drive the valley state to its ground state at the end of the shuttle.
Since this is a single-parameter optimization, it can be easily studied via an exhaustive search.
To this end, we study $10^3$ system configurations consisting of realizations of different valley landscapes $\vec{\Delta}(x)$ and initial angles $(\theta,\phi)$. 
To compare against the results obtained with the method proposed in the main text, we study shuttle velocities in the range of $10^2-10^3$m/s range.

\begin{figure}[h]
    \centering
    \includegraphics[width=\linewidth, trim={0cm 0cm 0cm 0cm},clip]{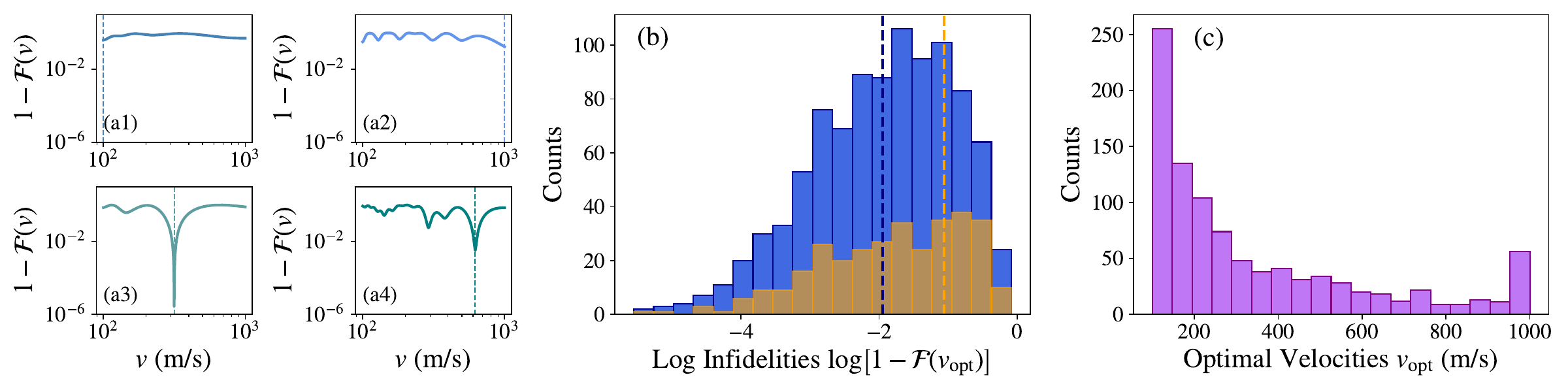}
    \caption{(a1–a4) Valley state infidelies as a function of shuttle velocity for four representative cases: (a1) slowest shuttle, (a2) fastest shuttle, (a3) shuttle yielding the smallest infidelity, and (a4) shuttle with infidelity near the median. 
    Vertical dashed lines represent the corresponding optimal velocities obtained from the search.
    (b) Histogram of optimal infidelities across $10^3$ system configurations. 
    The vertical dashed blue line indicates the median optimal infidelity of approximately $1-\mathcal{F}(v_\mathrm{opt})\approx10^{-2}$.
    In orange, we show the subset of infidelities for velocities smaller than $150$~m/s and larger than $950$~m/s, which has a mean infidelity of $\approx 0.09$ (orange dashed line), about 10 times larger than the total mean. 
    (c) Histogram of optimal shuttle velocities $v_\mathrm{opt}$ for the same configurations.}
    \label{fig:const_speed}
\end{figure}

For each configuration, we vary the shuttle velocity and compute the valley state infidelity at the end of the shuttle as the overlap with the ground state $\ket{\psi_-}$, as given by Eq.~\eqref{eq:F_V}.
Examples of infidelities as a function of shuttle velocities are shown in Fig.~\ref{fig:const_speed} for four representative cases.
Then, we record the optimum velocity that minimizes the state infidelity.
The statistics of the resulting optimized infidelities and optimal velocities are shown in Figs.~\ref{fig:const_speed}(b) and (c), respectively. 
In particular, a median infidelity of $10^{-2}$ was obtained. 
Although this value in itself is consistent with state-of-the-art shuttling accuracies, a significant fraction fall below the 99\% performance threshold, highlighting the limitations of the constant-speed approach.
Lastly, as shown in Fig.~\ref{fig:const_speed}(c), note that the largest density of shuttles are found near slower and higher velocities.
These cases, notably, present near an order of magnitude larger mean infidielities, as shown in orange in Fig.~\ref{fig:const_speed}(b).

An intuitive interpretation is that a single-velocity search changes both the total rotation angle and axis simultaneously, and consequently this method does not guarantee that the valley state can be controlled arbitrarily, as is required to drive the valley to the ground state during a fast shuttle. 
Our strategy introduced in this work, on the other hand, utilizes two velocities, and as we show in the main text it allows for sufficient flexibility of control in order to induce arbitrary valley rotations.

\section{Landau-Zener Instantaneous Eigenstates}
\label{app:valley_ground_state}

The standard Landau-Zener Hamiltonian in position space is defined by
\seq{
\label{eq:H_LZ_mat}
H_\mrm{LZ}(x) = \frac{1}{2} \begin{pmatrix}
\alpha x & g \\
g & -\alpha x
\end{pmatrix},
}
where $\alpha,g$ are assumed positive. 
This Hamiltonian can be diagonalized to find the instantaneous eigenstates. 
The eigenvalues are $E_\pm = \pm \frac{1}{2}\sqrt{g^2 + \alpha^2 x^2 }$.
In the same basis as $H_\mrm{LZ}$ in Eq.~\eqref{eq:H_LZ_mat}, the unnormalized eigenstates become
\seq{
\ket{\psi_\pm(x)} \dot{=} \begin{pmatrix}
g \\
\alpha x \pm \sqrt{\alpha^2 x^2 + g^2}
\end{pmatrix},
}
with $\ket{\psi_-}$ denoting the ground state with energy $E_-$, and the state is kept unnormalized to for simplicity.
At a distance $D$ away from the avoided crossing, for $g \ll \alpha D$, the eigenstate can be expanded to obtain
\seq{
\ket{\psi_-(D)} \approx \ket{0}-\frac{g}{2 \alpha D } \ket{1}.
}
Thus, we may write $\ket{\psi_-(D)} = \ket{0} + O(g/\alpha D)$.

\section{Approximating the Landau-Zener Evolution}

The first step of the optimization protocol is to linearize the intervalley coupling near the SLM located at $x_0$,
namely 
\seq{
H_\mrm{V}^\mrm{lin}(x)= \l( \l. \vec{\Delta}(x_0) + \frac{d\vec{\Delta}}{dx}\r|_{x_0}(x-x_0) \r) \cdot \frac{\vtau}{2},
}
where $x_0$ is defined such that $\l[\frac{d}{dx}|\Delta(x)|\r]_{x_0}=0$, from which the simple relationship between the phases of the coupling and its derivative can be derived, i.e., $\arg\Delta'(x_0)=\arg\Delta(x_0)+\pi/2$.
Next, to relate the linearized problem with a LZ transition, we define the unitary transformation 
\seq{
U_R = R_y(-\pi/2) R_x(\pi/2) R_z\left(\arg \Delta(x_0)\right),
}
where $R_k(\phi)=e^{-\frac{i}{2} \phi \tau_k}$.
Here, the initial $z$-rotation combines the position-dependencies of the two terms, followed by $x-$ and $y-$rotations to relabel the axes to match the standard LZ form.
Implementing this rotation onto $H_\mrm{V}^\mrm{lin}(x)$, it simplifies to the traditional LZ Hamiltonian
\seq{
\label{eq:app:H_LZ}
H_\mrm{LZ}(x) &= U_R H_\mrm{V}^\mrm{lin}(x) U_R^\dag \\
&= \frac{\alpha}{2}(x - x_0) \tau_z + \frac{g}{2} \tau_x,
}
where we defined the overall energy gap rate in space at the SLM as $\alpha=\left|\Delta'(x_0)\right|$ and the coupling magnitude as $g=|\Delta(x_0)|$, where the relationship between the phases of the coupling and its derivative was used explicitly.
An additional $R_z(\pi)$ rotation may be necessary to ensure $g>0$. 

Furthermore, from Eq.~\eqref{eq:app:H_LZ} we define the characteristic distance $d \equiv g/\alpha$.
Note that, since $H_\mrm{LZ}(x)$ is approximately proportional to $\tau_z$ for $|x-x_0|\gg d$, the non-trivial contribution to the valley evolution will occur in the region near the SLM. 
As will be shown below, this allows us to treat the electron’s passage through regions of small valley splitting as a controlled LZ transition by tuning the velocity profile.
Next, we analytically solve $H_\mrm{V}^\mrm{lin}(x)$ to obtain a good seed pulse from which to numerically optimize the actual evolution due to $H_\mrm{V}(x)$.

\begin{figure}[h]
    \centering
    \includegraphics[width=.7\linewidth, trim={0cm 7.5cm 9cm 0cm},clip]{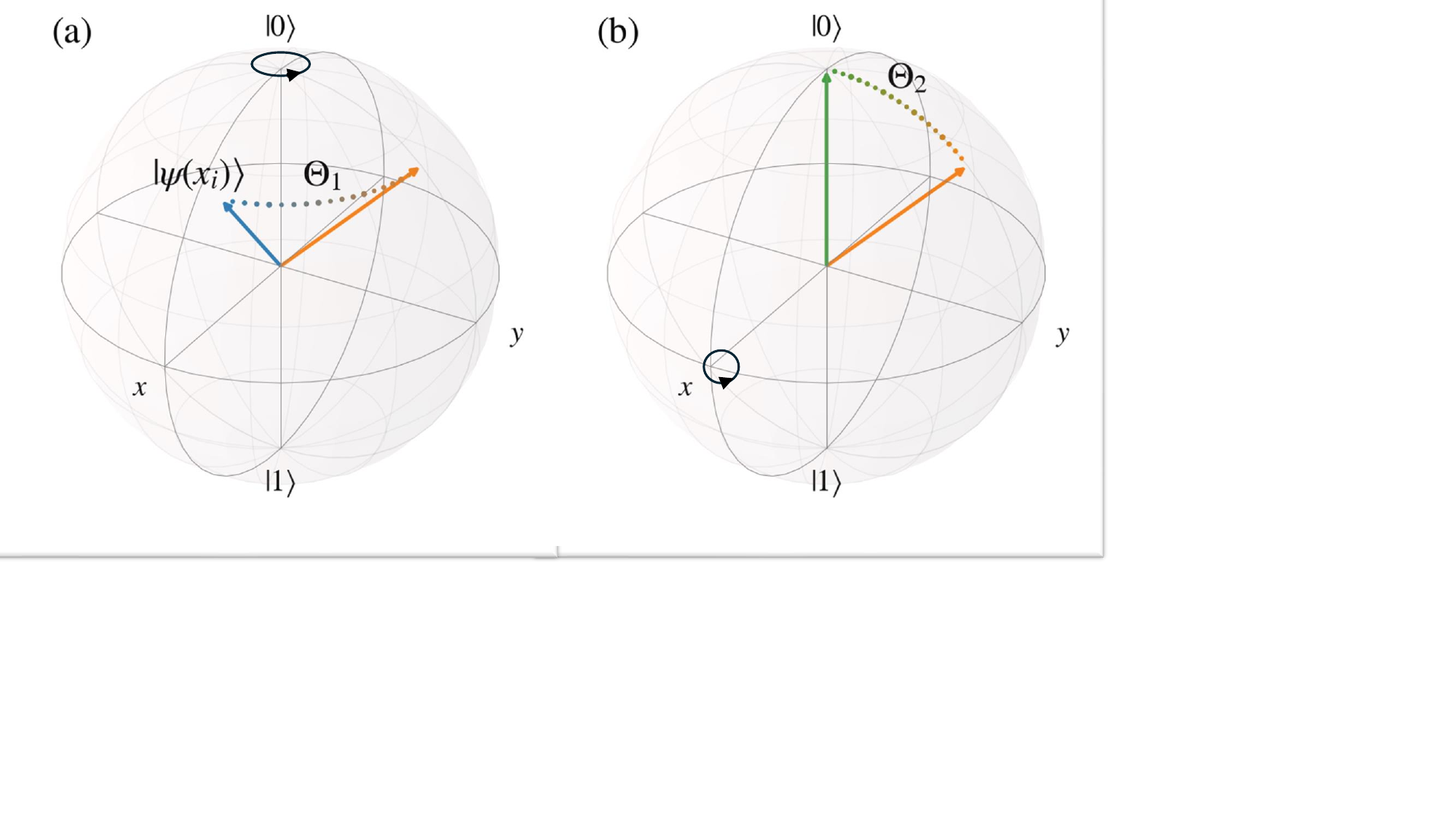}
    \caption{Composition of rotations in the PEA, with blue vector corresponding to the initial state $\ket{\psi(x_i)}$ in the LZ frame. (a) $z$-rotation bringing the initial state $\ket{\psi(x_i)}$ to the $yz$-plane, where the orange vector corresponds to the state $R_z\l(\Theta_1\r)\ket{\psi(x_i)}$. (b) $x$-rotation bringing the intermediate state to the $\ket{0}$ state, where the green vector corresponds to the state $ R_x\l(\Theta_2\r) R_z\l(\Theta_1\r)\ket{\psi(x_i)}$.}
    \label{fig:bloch}
\end{figure}

We approximate the total LZ evolution as a composition of three rotations, $z \rightarrow x \rightarrow z$, corresponding to the electron starting far from the SLM ($|x_0-x| \gg d$) where the $\tau_z$ term dominates, then instantaneously being shuttled to the SLM ($x \approx x_0$) where the $\tau_x$ term dominates, and finally instantaneously moving far away again ($|x-x_0| \gg d$).
We will refer to this as the piecewise evolution approximation (PEA), where the corresponding Hamiltonian that generates this evolution is
\seq{
\label{eq:app:H_PEA}
H_\mrm{PEA}(x) = \frac{1}{2}
\begin{cases}
\alpha\, (x-x_0) \, \tau_z & \mrm{if} \quad |x-x_0|>d,\\
g \,\tau_x & \mrm{if} \quad |x-x_0|<d
\end{cases}.
}
The total propagator $U_\mrm{V}(t_f;t_i)$ is then approximated by
\begin{equation}
\label{eq:U_PEA}
U_\mrm{PEA} = e^{-\frac{i}{2} \Theta_3 \tau_z} e^{- \frac{i}{2} \Theta_2 \tau_x} e^{-\frac{i}{2} \Theta_1 \tau_z},
\end{equation}
where $\l\{\Theta_j\r\}_{j=1,2,3}$ correspond to the effective rotation angles in each region, written in terms of LZ Hamiltonian parameters as
\seq{
\label{eq:app:PEA_angles_def}
\Theta_1 &= \alpha \l[\int_{t_i}^{t_0-\tau} x(t) dt + x_0(t_i-t_0 + \tau) \r], \\
\Theta_2 &= g \int_{t_0-\tau}^{t_0+\tau} dt = 2 g \tau , \\
\Theta_3 &= \alpha \l[\int_{t_0+\tau}^{t_f} x(t) dt  - x_0(t_f -t_0 - \tau)\r].
}
Here we defined the times $t_0$ and $\tau$ such that $x(t_0)=x_0$ and $x(t_0\pm\tau)=x_0\pm d$.
By choosing $x(t)$ appropriately and with sufficiently long shuttling times, any rotations can be achieved.
Lastly, $U_\mrm{PEA}$ can be interpreted as an Euler decomposition of a general unitary~\cite{McKay2017_virtual}, thus enabling the treatment of the PEA evolution as a control problem.
An example of this evolution is shown in Fig.~\ref{fig:bloch}.

As the electron moves away from the SLM, the instantaneous ground state $\ket{0}$ approaches the valley ground state, $\ket{0}= \ket{\psi_-}+\mathcal{O}(d/D)$.
Since in cases of interest $D \gg d$, we use $\ket{0}$ as the PEA target state.
To find the transformation of interest, we first parameterize the initial valley state at the starting location $x_i$ in standard Bloch sphere representation~\cite{Nielsen2012},
\begin{equation}
\ket{\psi(x_i)} = \cos(\theta_i/2) \ket{0} + e^{i\phi_i} \sin(\theta_i/2) \ket{1}.
\end{equation}
Then, we find the Bloch angles $(\phi,\theta)$ in the LZ frame by computing $U_R\ket{\psi(x_i)}=\cos(\theta/2) \ket{0} + e^{i\phi} \sin(\theta/2) \ket{1}$.
From Eq.~\eqref{eq:U_PEA}, $U_\mrm{PEA} \ket{\psi(x_i)}=\ket{0}$ if
\seq{\label{eq:Theta}
\Theta_1 &= \frac{\pi}{2} - \phi + 2 \pi n_1, \\
\Theta_2 &= \theta + 2 \pi n_2,
}
where the choice of integer $n_1 (n_2)$ correspond to the number of complete circuits rotated about the $z(x)-$axis. 
The choice of $\Theta_3$ is arbitrary.

Next, we find the required physical parameters in the LZ frame from these rotation angles and Eqs.~\eqref{eq:app:PEA_angles_def}.
Following the scheme suggested by the PEA in Eq.~\eqref{eq:app:H_PEA}, we segment the shuttling process into three regions of constant velocity, namely
\seq{\label{eq:app:v}
v(t) = \begin{cases}
    v_1 & \mathrm{if}\quad |t-t_0| > \tau,\\
    v_2 & \mathrm{if}\quad |t-t_0| < \tau,
\end{cases}
}
where we imposed a symmetric condition on the velocities for simplicity.
Here, we denote $(t_0-\tau,t_0+\tau)$ the range of times over which the electron is expected to be near the SLM.
Denoting $T = t_0 - t_i = (D-d)/v_1 + \tau$, integrating the first line of Eq.~\eqref{eq:app:PEA_angles_def} then results in
\seq{
\Theta_1 &= \alpha \l[ \frac{v_1}{2} (T-\tau)^2 - D (T- \tau) \r].
}
Noting that $\tau = d/v_2=g/(\alpha v_2)$ and combining with Eq.~\eqref{eq:Theta} yields
$\tau = \frac{\theta + 2 \pi n_2}{2 g}$
and the corresponding shuttling velocities in the PEA are
\seq{
\label{eq:v_PEA}
v_1 &= \frac{\alpha\l( D^2 - d^2\r)}{2\phi-\pi (4 n_1+1)}, \\
v_2 &= \frac{2 g^2}{\alpha (\theta + 2 \pi n_2)}.
}
Lastly, the value of $t_f$ is set by imposing $x(t_f)=L$, namely $t_f=t_0+\tau+(L - x_0 - d)/v_1$. See next section for details on optimal choices of $n_1,n_2$.

\section{Optimal Choices of \texorpdfstring{$n_1,n_2,T$}{}}
\label{app:opt_n1n2}

As shown in Eq.~\eqref{eq:v_PEA}, there exist multiple combinations of $n_1,n_2$ that meet the conditions required to shuttle the electron while suppressing valley-induced dephasing. 
Here we propose a criterion to choose $n_1,n_2$ based on efficiency under physical constraints.

First, a relationship can be derived for $n_2$ by enforcing $\tau>0$ and requiring $\tau$ to minimize the shuttling time under the maximum shuttling velocity constraint $|v(t)|\lesssim v_\mrm{max}$.
From imposing this condition on $v_2^\mrm{PEA}$, we obtain
\seq{
n_2^\mrm{opt} = \l\lceil \frac{1}{2\pi} \l( \frac{2 g^2}{\alpha v_\mrm{max}}-\theta\r) \r\rceil,
}
from where the minimum $\tau_\mrm{min}=(\theta+2\pi n_2^\mrm{opt})/g$ can be found. In practice, since $v_\mrm{max}=10^3\,$m/s and $g\ll \alpha D$, this implies that $n_2$ is set to small values, namely 0 or $-1$.

Next, to ensure that the maximum velocity condition is met, we impose $T_\mrm{min} \gtrsim \tau_\mrm{min} + (D-d)/v_\mrm{max}$. 
During the optimization, the optimizer may fail to achieve sufficiently high fidelities. When this is the case, we iterate over $T$ in the range $(T_\mrm{min},2T_\mrm{min})$ and find new PEA parameters. See main text for details of this performance for iterations over $T$.

Lastly, for the PEA to approximate the physical valley problem, the traversed distance near the avoided crossing $2 v_2 \tau$ must be close to the transition distance $2 d$. 
Defining the LZ energy a distance $D$ away from the avoided crossing $\Delta_i= \alpha D$, we write $d$ in terms of the physical and optimal parameters,
\seq{
v_2 \tau &= D - v_1 (T-\tau)\\
&= \frac{1}{\alpha}\l( \Delta_i - \l[ \frac{\pi(4n_1 +1)-2\phi}{(T-\tau)^2}+ \frac{2\Delta_i}{T-\tau} \r] (T-\tau) \r) \\
&= \frac{1}{\alpha}\l( - \Delta_i + \frac{2\phi-(4 n_1 + 1 )\pi}{T-\tau}\r)
}
which, from requiring $v_2 \tau \approx g/\alpha$,
implies $(4n_1+1)\pi = 2\phi-(T-\tau)(g+\Delta_i)$. 
Consequently,  we obtain
\eq{
n_1^\mrm{opt} = \l\lceil \frac{2\phi-\pi-(T-\tau)(g+\Delta_i)}{4\pi} \r\rceil.
}
Thus, taking these values of $n_1,n_2,T$ we ensure that the PEA solution is achieved while approximating the real evolution, and at the same time imposing the physical constraint of maximum shuttling speed but maintaining shuttling efficiency.

\section{Valley-Informed Optimization}

When detailed information of the valley landscape is accessible, it is possible to differentiate the Hamiltonian explicitly, and thus execute a gradient-based optimization instead of the gradient-free approach introduced in the main text. 
Since the location of the SLM is known, only the velocities must be optimized.
Using the ansatz of Eq.~\eqref{eq:v}, we optimize only over $(v_1,v_2)$, while the values of $t_0,\tau$ are obtained from the PEA.
From here on we will denote this type of optimization as valley-informed (VI), whereas we will refer to the one introduced in the main text by valley-agnostic (VA).

\begin{figure*}[h]
    \centering
    \includegraphics[width=0.75\linewidth, trim={0cm .45cm 0cm 0cm},clip]{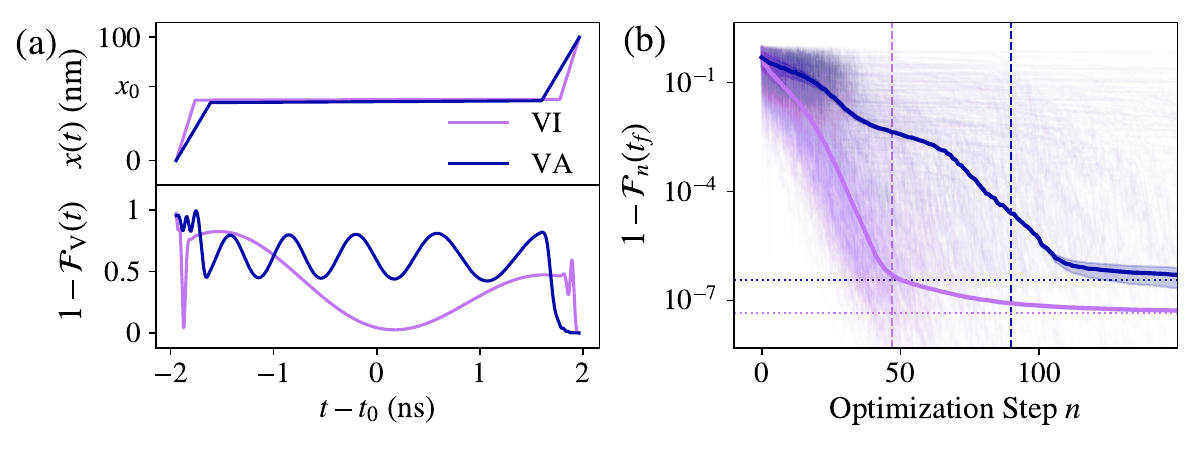}
    \caption{VI optimization results (purple); VA results (blue) are repeated from Fig.~\ref{fig:results} to aid comparison (see main text for optimization specifics). (a) Example corresponding to shuttling past the last SLM shown in Fig.~\ref{fig:main}. (b) Convergence properties.
    }
    \label{fig:results_VIVA}
\end{figure*}

A representative example of the valley fidelity obtained from optimization can be found in Fig.~\ref{fig:results_VIVA}(a) for the valley landscape near the last SLM shown in Fig.~\ref{fig:main}, where we show the results of VA optimization corresponding to Fig.~\ref{fig:results}(a) of the main text.
As was the case for VA, the electron spends most of the shuttling around the SLM as now VI shuttles quickly to that point at $\sim750\,$m/s, while around the SLM dropping to within the $\pm 1\,$m/s range.
We tested the VI optimization on the same set of landscapes and parameters used in the main text, and found that the VI optimization needs approximately half the number of iterations to achieve an order of magnitude lower infidelity. The results are shown in Figs.~\ref{fig:results_VIVA}(b) and (c), and summarized in Table~\ref{tab:table_VI_VA}.

\begin{table}[h]
    \centering
    \begin{tabular}{||c|c|c||}
    \hline
        Median & VI & VA \\
         \hline
        Steps to convergence & 47 & 90 \\
        Noiseless infidelity & $4\times 10^{-8}$ & $4\times 10^{-7}$ \\
        Noisy infidelity & $\approx p_\mrm{err}$ & $\approx p_\mrm{err}$\\
        \hline
    \end{tabular}
    \caption{Performance comparison between VI and VA approaches summarizing the results in Fig.~\ref{fig:results_VIVA}. VI provides advantages both in convergence speed and optimized infidelities. The noisy infidelities converge to measurement noise $p_\mrm{err}$ for a sufficiently large number of shots ($M\gtrsim 100$). }
    \label{tab:table_VI_VA}
\end{table}

\section{Optimization Details}
\label{app:landscapes}

\begin{figure}[h]
    \centering
    \includegraphics[width=\linewidth, trim={0cm 0cm 0cm 0cm},clip]{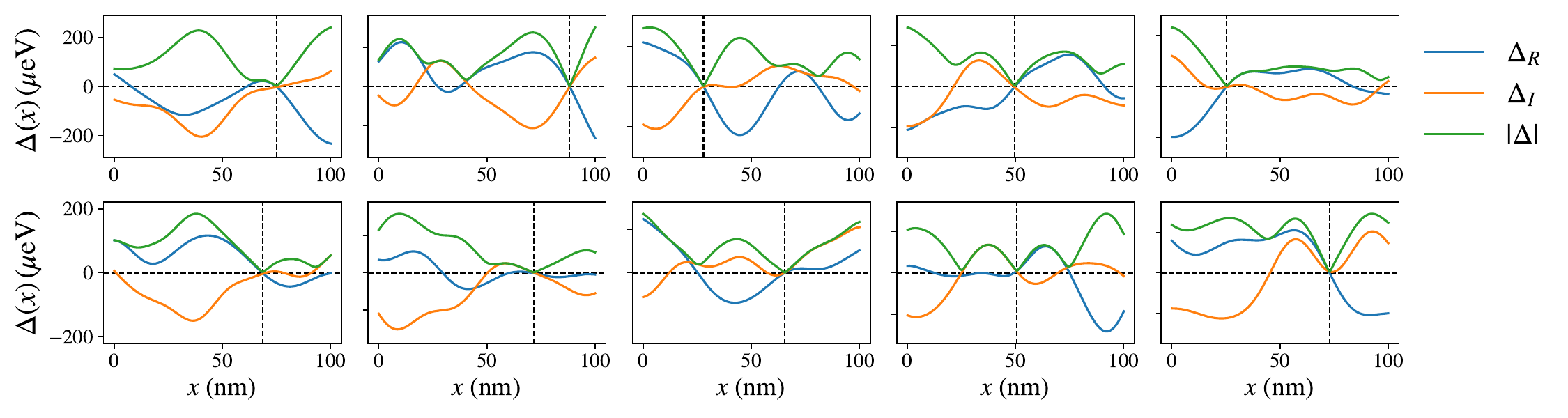}
    \caption{Representative examples of randomly generated valley landscapes used in the main text to test the optimization protocols.
    The landscapes were generated following the steps outlined in Ref.~\cite{Losert2024Strategies}. 
    Vertical lines show the location of chosen SLMs around which the velocity profiles are optimized for each landscape.}
    \label{fig:landscapes}
\end{figure}

In this section we provide the details used in the numerical optimizations performed in the main text. 
For the generation of valley landscapes, we followed the steps outlined in Ref.~\cite{Losert2024Strategies}
, where the authors introduced a simple way of generating realistic landscapes using a Gaussian covariance model and the Python package GSTools to generate fluctuating Si concentrations.
Here, we define a symmetric dot size of $a_x = a_y = 14\,$nm, which sets the correlation lengths of $\Delta_R$ and $\Delta_I$.
From this dot size, we chose a shuttling distance of $100\,$nm in order for at least one SLM to be present. 
The covariance $\sigma_\Delta^2$ of the coupling $\Delta$ was set to $\sigma_\Delta = 100\,\mu$eV, to match order of magnitudes observed in experimental measurements.
From here, it can be derived that the covariances of $\Delta_R$ and $\Delta_I$ are given by $\sigma_\Delta^2 / 2$.
The average valley splitting used to generate these landscapes is $\sqrt{\pi}/2 \sigma_\Delta$.
We assume the dot is the ground state of an isotropic harmonic oscillator potential with mass $m_t$ and level spacing $\hbar \omega = 2\,$meV, giving rise to dot radii $a_x = a_y = \sqrt{\hbar/(m_t \omega)} = 14\,$nm.
Representative examples are shown in Fig.~\ref{fig:landscapes}.
Notice that some landscapes can be interpreted as presenting more than one SLM in this range, which, as explained in the main text, does not constitute a problem for our method.
When multiple SLM are present, the optimization is performed over the one presenting the minimum $|\Delta(x)|$.

The 100 initial angles $(\phi_V,\theta_V)$ were chosen uniformly from the Bloch sphere using the Fibonacci lattice method~\cite{Swinbank2006,Gonzlez2009}. 
The north and south poles of the sphere, corresponding to the computational basis states, were included.

VA optimization initial seeds were drawn from normal distributions based on the mean and standard deviations of the parameters obtained from the VI optimization results. The mean and standard deviation values found from the $10^4$ VI optimization are: $v_1 \in (300 \pm 100)\,\mrm{m/s},\, v_2 \in (0.1 \pm 2)\,\mrm{m/s},\,\tau\in(1.5\pm 2)\,\mrm{ns}$.
The VA seeds were drawn using the Python package NumPy's random.normal functionality, with center and scale parameters set to the mean and standard deviations obtained from VI, respectively.
Note that if a given value of $\tau<0$ is drawn, a new value is sampled until $\tau>0$ is obtained. 
Lastly, the parameter $t_0$ is chosen to locate the middle region in the center of the shuttle, i.e., $t_0=\tau + (L/2-v_2 \tau)/v_1$.

\section{Optimization Success Rate}
\label{app:success_rate}

In this section we study the success rate of the optimizations for given initial conditions.
We define the success rate in optimization as the fraction of optimizations that achieve an infidelity below $10^{-4}$.
This threshold is chosen to be representative of state-of-the-art gate infidelities.
The protocol employed involves first optimizing the entire set of $10^4$ different test cases, and then selectively reoptimizing only the cases that fail to meet the specified fidelity threshold. 
For VI, the attempts refer to extending the first shuttling time by increasing $T$ in the range $(T_\mrm{min},2 T_\mrm{min})$ in steps of $0.1 T_\mrm{min}$.
In the VA case, new initial conditions are drawn with each attempt.
See Sec.~\ref{app:opt_n1n2} for details on the choice of $T_\mrm{min}$ and the initial condition sampling of VA.

\begin{figure}[h]
    \centering
    \includegraphics[width=.4\linewidth]{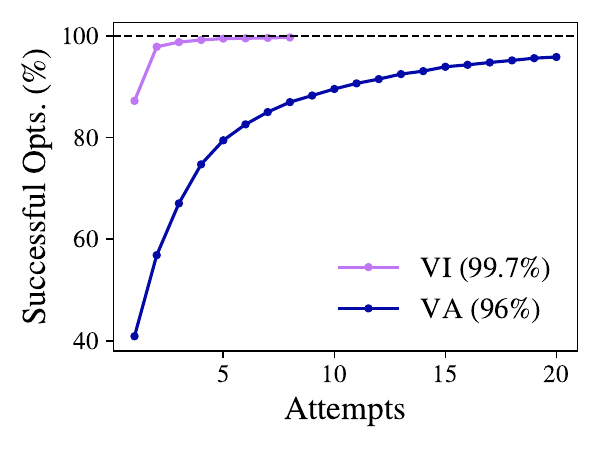}
    \caption{Comparison of optimization success rates between VI and VA. VI achieves a 99\% success rate after 3 attempts, while VA reaches 90\% after 10 attempts. In the plot legend we present the maximum success rates obtained after the number of attempts shown.}
    \label{fig:success_rate}
\end{figure}

Figure~\ref{fig:success_rate} shows the success rates after each iteration for both methods.
As shown, the VI method obtains a success rate of 88\% after the first iteration, quickly achieving 99\%.
The VA method on the other hand, is successful on 40\% of the test cases, while increasing to 90\% after 10 iterations. 
this highlights the fact that knowledge of the valley landscape provided by the PEA initial seed in the VI case greatly improves the optimization performance.
However, this also provides further support to detailed valley landscape information not being essential for the success of the optimization.

\section{Quantifying Noise Induced by Signal Uncertainties and Charge Noise}

To evaluate the robustness of the protocol to control imperfections, we simulate the effect of noise on both the optimized velocity parameters $(v_1,v_2)$ and the timing parameters $(t_0,\tau)$. Noise is introduced independently for each run, using two different models: multiplicative noise for velocities and additive noise for timings. Specifically, the velocities are modified as $v_i \rightarrow v_i(1+\xi_i)$, where $\xi_i$ is drawn from a zero-mean Gaussian distribution with standard deviation $\xi$. Timing jitter is modeled as $t_0 \rightarrow t_0+\delta_{t_0}$ and $\tau \rightarrow \tau+\delta_\tau$, with $\delta_{t_0,\tau}$ also drawn from a zero-mean Gaussian distribution, now with standard deviation $\delta t$. While reported experimental jitter values are typically in the range of 2–4~ps~\cite{Mao2022,Wang2024}, we extend the simulated noise range up to 0.1~ns to reflect the AWG resolution and to probe the operational margin.

In addition to these signal-based uncertainties, we also account for charge-noise–induced displacements of the quantum dot position. Near regions of small valley splitting, even sub-nanometer fluctuations in dot location can shift the effective valley splitting and thereby interfere with the optimized transfer protocol. To capture this effect, we model rigid shifts of the dot position around the SLM location $x_0$ as $\Delta(x)\rightarrow \Delta(x+\delta_x)$, where $\delta_x$ is a zero-mean Gaussian random variable with standard deviation $\delta x$. Experimental estimates place such fluctuations in the order of $\sim 10$~pm~\cite{Yoneda2017}, which we use as a reference scale in our analysis.

\begin{figure}[h]
    \centering
    \includegraphics[width=\linewidth]{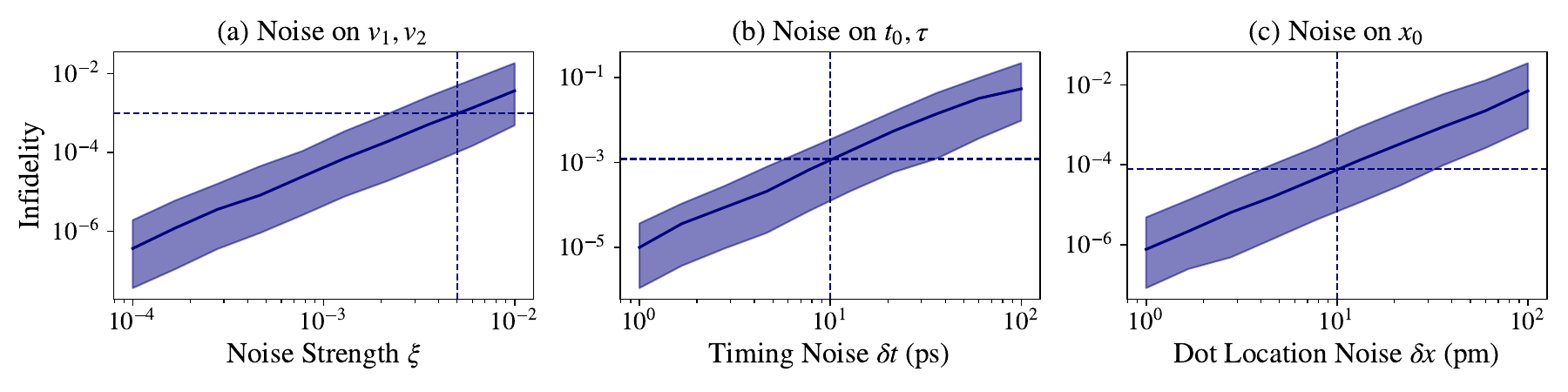}
    \caption{Average valley state infidelities as function of: (a) multiplicative noise strength with average $\xi$ applied on shuttle velocities, (b) additive timing noise on the switching times given by $t_0,\tau$, as function of noise amplitude $\delta t$, and (c) on location of the dot, with amplitude $\delta x$. Each curve shows the median infidelity over 100 noise realizations, computed across 15 different disorder landscapes and initial condition configurations. Shaded bands represent the interquartile ranges. Vertical and horizontal dotted lines mark reference noise strengths of: $\xi=5 \cdot 10^{-3}$, $\delta t=10~$ps and $\delta x=10~$pm, and their corresponding average infidelity values, respectively. Note that these are compatible with infidelities in the range of $10^{-4}$--$10^{-3}$.}
    \label{fig:signal_uncertainties}
\end{figure}

Figure~\ref{fig:signal_uncertainties} summarizes the results of these simulations. For each noise amplitude, we calculate the median infidelity (as defined in the main text) over 100 noisy realizations, sampled across 15 disorder landscapes and initial conditions. The results show that fidelities remain below $10^{-3}$ for velocity noise strengths $\xi \leq 5 \cdot 10^{-3}$ and for timing noise $\delta t \leq 10$~ps, and below $10^{-4}$ for dot displacement noise $\delta x \leq 10$~pm. Importantly, in all cases the infidelity increases smoothly with noise amplitude, indicating that the optimized sequences exhibit no sharp instabilities even when designed without explicit awareness of these uncertainties.

\newpage

\putbib[references]
\end{bibunit}


\begin{thebibliography}{50}%
\makeatletter
\providecommand \@ifxundefined [1]{%
 \@ifx{#1\undefined}
}%
\providecommand \@ifnum [1]{%
 \ifnum #1\expandafter \@firstoftwo
 \else \expandafter \@secondoftwo
 \fi
}%
\providecommand \@ifx [1]{%
 \ifx #1\expandafter \@firstoftwo
 \else \expandafter \@secondoftwo
 \fi
}%
\providecommand \natexlab [1]{#1}%
\providecommand \enquote  [1]{``#1''}%
\providecommand \bibnamefont  [1]{#1}%
\providecommand \bibfnamefont [1]{#1}%
\providecommand \citenamefont [1]{#1}%
\providecommand \href@noop [0]{\@secondoftwo}%
\providecommand \href [0]{\begingroup \@sanitize@url \@href}%
\providecommand \@href[1]{\@@startlink{#1}\@@href}%
\providecommand \@@href[1]{\endgroup#1\@@endlink}%
\providecommand \@sanitize@url [0]{\catcode `\\12\catcode `\$12\catcode
  `\&12\catcode `\#12\catcode `\^12\catcode `\_12\catcode `\%12\relax}%
\providecommand \@@startlink[1]{}%
\providecommand \@@endlink[0]{}%
\providecommand \url  [0]{\begingroup\@sanitize@url \@url }%
\providecommand \@url [1]{\endgroup\@href {#1}{\urlprefix }}%
\providecommand \urlprefix  [0]{URL }%
\providecommand \Eprint [0]{\href }%
\providecommand \doibase [0]{http://dx.doi.org/}%
\providecommand \selectlanguage [0]{\@gobble}%
\providecommand \bibinfo  [0]{\@secondoftwo}%
\providecommand \bibfield  [0]{\@secondoftwo}%
\providecommand \translation [1]{[#1]}%
\providecommand \BibitemOpen [0]{}%
\providecommand \bibitemStop [0]{}%
\providecommand \bibitemNoStop [0]{.\EOS\space}%
\providecommand \EOS [0]{\spacefactor3000\relax}%
\providecommand \BibitemShut  [1]{\csname bibitem#1\endcsname}%
\let\auto@bib@innerbib\@empty

\bibitem{Fujita2017Coherent}
  T. Fujita, T. A. Baart, C. Reichl, W. Wegscheider, and L. M. K. Vandersypen, npj Quantum Inf. \textbf{3}, 22 (2017).

\bibitem{Mills2019Shuttling}
  A. R. Mills, D. M. Zajac, M. J. Gullans, F. J. Schupp, T. M. Hazard, and J. R. Petta, Nat. Commun. \textbf{10}, 1063 (2019).

\bibitem{Yoneda2021Coherent}
  J. Yoneda, W. Huang, M. Feng, C. H. Yang, K. W. Chan, T. Tanttu, W. Gilbert, R. C. C. Leon, F. E. Hudson, K. M. Itoh, A. Morello, S. D. Bartlett, A. Laucht, A. Saraiva, and A. S. Dzurak, Nat. Commun. \textbf{12}, 4114 (2021).

\bibitem{Jadot2021Distant}
  B. Jadot, P.-A. Mortemousque, E. Chanrion, V. Thiney, A. Ludwig, A. D. Wieck, M. Urdampilleta, C. B\"auerle, and T. Meunier, Nat. Nanotechnol. \textbf{16}, 570 (2021).

\bibitem{Noiri2022shuttling}
  A. Noiri, K. Takeda, T. Nakajima, T. Kobayashi, A. Sammak, G. Scappucci, and S. Tarucha, Nat. Commun. \textbf{13}, 5740 (2022).

\bibitem{Boter2022Spiderweb}
  J. M. Boter, J. P. Dehollain, J. P. van Dijk, Y. Xu, T. Hensgens, R. Versluis, H. W. Naus, J. S. Clarke, M. Veldhorst, F. Sebastiano, and L. M. Vandersypen, Phys. Rev. Appl. \textbf{18}, 024053 (2022).

\bibitem{Zwerver2023Shuttling}
  A. Zwerver, S. Amitonov, S. de Snoo, M. Madzik, M. Rimbach-Russ, A. Sammak, G. Scappucci, and L. Vandersypen, PRX Quantum \textbf{4}, 030303 (2023).

\bibitem{Struck2024Spin}
  T. Struck, M. Volmer, L. Visser, T. Offermann, R. Xue, J.-S. Tu, S. Trellenkamp, L. Cywinski, H. Bluhm, and L. R. Schreiber, Nat. Commun. \textbf{15}, 1325 (2024).

\bibitem{Tyryshkin2006Coherence}
  A. M. Tyryshkin, J. J. L. Morton, S. C. Benjamin, A. Ardavan, G. A. D. Briggs, J. W. Ager, and S. A. Lyon, J. Phys. Condens. Matter \textbf{18}, S783 (2006).

\bibitem{Burkard2023Semiconductor}
  G. Burkard, T. D. Ladd, A. Pan, J. M. Nichol, and J. R. Petta, Rev. Mod. Phys. \textbf{95}, 025003 (2023).

\bibitem{Harvey-Collard2022Resonator}
  P. Harvey-Collard, J. Dijkema, G. Zheng, A. Sammak, G. Scappucci, and L. M. K. Vandersypen, Phys. Rev. X \textbf{12}, 021026 (2022).

\bibitem{Seidler2022Conveyor}
  I. Seidler, T. Struck, R. Xue, N. Focke, S. Trellenkamp, H. Bluhm, and L. R. Schreiber, npj Quantum Inf. \textbf{8}, 100 (2022).

\bibitem{Langrock2023Blueprint}
  V. Langrock, J. A. Krzywda, N. Focke, I. Seidler, L. R. Schreiber, and L. Cywi\'nski, PRX Quantum \textbf{4}, 020305 (2023).

\bibitem{Xue2024Si}
  R. Xue, M. Beer, I. Seidler, S. Humpohl, J.-S. Tu, S. Trellenkamp, T. Struck, H. Bluhm, and L. R. Schreiber, Nat. Commun. \textbf{15}, 2296 (2024).

\bibitem{white2024electricalinterconnectssiliconspin}
  C. D. White, A. Sigillito, and M. J. Gullans, \href{https://arxiv.org/abs/2411.01366}{arXiv:2411.01366}.

\bibitem{Zwanenburg2013Silicon}
  F. A. Zwanenburg, A. S. Dzurak, A. Morello, M. Y. Simmons, L. C. L. Hollenberg, G. Klimeck, S. Rogge, S. N. Coppersmith, and M. A. Eriksson, Rev. Mod. Phys. \textbf{85}, 961 (2013).

\bibitem{Friesen2007}
  M. Friesen, S. Chutia, C. Tahan, and S. N. Coppersmith, Phys. Rev. B \textbf{75}, 115318 (2007).

\bibitem{Culcer2010}
  D. Culcer, X. Hu, and S. Das Sarma, Phys. Rev. B \textbf{82}, 205315 (2010).

\bibitem{Mi2018}
  X. Mi, S. Kohler, and J. R. Petta, Phys. Rev. B \textbf{98}, 161404 (2018).

\bibitem{Lima2024Valley}
  J. R. F. Lima and G. Burkard, Phys. Rev. Mater. \textbf{8}, 036202 (2024).

\bibitem{Tariq2019Effects}
  B. Tariq and X. Hu, Phys. Rev. B \textbf{100}, 125309 (2019).

\bibitem{Gamble2013Disorder}
  J. K. Gamble, M. A. Eriksson, S. N. Coppersmith, and M. Friesen, Phys. Rev. B \textbf{88}, 035310 (2013).

\bibitem{Ruskov2018Electron}
  R. Ruskov, M. Veldhorst, A. S. Dzurak, and C. Tahan, Phys. Rev. B \textbf{98}, 245424 (2018).

\bibitem{Friesen2007Valley}
  M. Friesen, S. Chutia, C. Tahan, and S. N. Coppersmith, Phys. Rev. B \textbf{75} (2007).

\bibitem{DeSmet2024High}
  M. De Smet, Y. Matsumoto, A.-M. J. Zwerver, L. Tryputen, S. L. de Snoo, S. V. Amitonov, A. Sammak, N. Samkharadze, O. G\"ul, R. N. M. Wasserman, M. Rimbach-Russ, G. Scappucci, and L. M. K. Vandersypen, Nat. Nanotechnol. \textbf{20}, 866 (2025).

\bibitem{Losert2024Strategies}
  M. P. Losert, M. Oberl\"ander, J. D. Teske, M. Volmer, L. R. Schreiber, H. Bluhm, S. Coppersmith, and M. Friesen, PRX Quantum \textbf{5}, 040322 (2024).

\bibitem{Hwang2017Impact}
  J. C. C. Hwang, C. H. Yang, M. Veldhorst, N. Hendrickx, M. A. Fogarty, W. Huang, F. E. Hudson, A. Morello, and A. S. Dzurak, Phys. Rev. B \textbf{96}, 045302 (2017).

\bibitem{Feng2023Control}
  M. Feng, J. Yoneda, W. Huang, Y. Su, T. Tanttu, C. H. Yang, J. D. Cifuentes, K. W. Chan, W. Gilbert, R. C. C. Leon, F. E. Hudson, K. M. Itoh, A. Laucht, A. S. Dzurak, and A. Saraiva, Phys. Rev. B \textbf{107}, 085427 (2023).

\bibitem{Losert2023Practical}
  M. P. Losert, M. A. Eriksson, R. Joynt, R. Rahman, G. Scappucci, S. N. Coppersmith, and M. Friesen, Phys. Rev. B \textbf{108}, 125405 (2023).

\bibitem{Lima2024Superadiabatic}
  J. R. F. Lima and G. Burkard, Phys. Rev. B \textbf{111}, 235439 (2025).

\bibitem{Ivakhnenko2023Nonadiabatic}
  O. V. Ivakhnenko, S. N. Shevchenko, and F. Nori, Phys. Rep. \textbf{995}, 1 (2023).

\bibitem{Vitanov1996LandauZener}
  N. V. Vitanov and B. M. Garraway, Phys. Rev. A \textbf{53}, 4288 (1996).

\bibitem{Volmer2024Mapping}
  M. Volmer, T. Struck, A. Sala, B. Chen, M. Oberl\"ander, T. Offermann, R. Xue, L. Visser, J.-S. Tu, S. Trellenkamp, L. Cywinski, H. Bluhm, and L. R. Schreiber, npj Quantum Inf. \textbf{10}, 61 (2024).

\bibitem{SM}
  See Supplemental Material at \url{http://link.aps.org/supplemental/10.1103/s624-dvvs} for supporting calculations.

\bibitem{Kayanuma1997}
  Y. Kayanuma, Phys. Rev. A \textbf{55}, R2495 (1997).

\bibitem{Nelder1965}
  J. A. Nelder and R. Mead, Comput. J. \textbf{7}, 308 (1965).

\bibitem{McKay2017_virtual}
  D. C. McKay, C. J. Wood, S. Sheldon, J. M. Chow, and J. M. Gambetta, Phys. Rev. A \textbf{96}, 022330 (2017).

\bibitem{Johansson2013_qutip}
  J. Johansson, P. Nation, and F. Nori, Comput. Phys. Commun. \textbf{184}, 1234 (2013).

\bibitem{Yang2013}
  C. H. Yang, A. Rossi, R. Ruskov, N. S. Lai, F. A. Mohiyaddin, S. Lee, C. Tahan, G. Klimeck, A. Morello, and A. S. Dzurak, Nat. Commun. \textbf{4}, 2069 (2013).

\bibitem{Connors2022}
  E. J. Connors, J. Nelson, L. F. Edge, and J. M. Nichol, Nat. Commun. \textbf{13}, 940 (2022).

\bibitem{ye2024characterizationindividualchargefluctuators}
  F. Ye, A. Ellaboudy, D. Albrecht, R. Vudatha, N. T. Jacobson, and J. M. Nichol, Phys. Rev. B \textbf{110}, 235305 (2024).

\bibitem{Yoneda2023}
  J. Yoneda, J. S. Rojas-Arias, P. Stano, K. Takeda, A. Noiri, T. Nakajima, D. Loss, and S. Tarucha, Nat. Phys. \textbf{19}, 1793 (2023).

\bibitem{Jock2022}
  R. M. Jock, N. T. Jacobson, M. Rudolph, D. R. Ward, M. S. Carroll, and D. R. Luhman, Nat. Commun. \textbf{13}, 641 (2022).

\bibitem{Yoneda2017}
  J. Yoneda, K. Takeda, T. Otsuka, T. Nakajima, M. R. Delbecq, G. Allison, T. Honda, T. Kodera, S. Oda, Y. Hoshi, N. Usami, K. M. Itoh, and S. Tarucha, Nat. Nanotechnol. \textbf{13}, 102 (2017).

\bibitem{MacQuarrie2020}
  E. R. MacQuarrie, S. F. Neyens, J. P. Dodson, J. Corrigan, B. Thorgrimsson, N. Holman, M. Palma, L. F. Edge, M. Friesen, S. N. Coppersmith, and M. A. Eriksson, npj Quantum Inf. \textbf{6}, 81 (2020).

\bibitem{Gorman2005}
  J. Gorman, D. G. Hasko, and D. A. Williams, Phys. Rev. Lett. \textbf{95}, 090502 (2005).

\bibitem{Mi2017}
  X. Mi, J. V. Cady, D. M. Zajac, P. W. Deelman, and J. R. Petta, Science \textbf{355}, 156 (2017).

\bibitem{Elzerman2004}
  J. M. Elzerman, R. Hanson, L. H. Willems van Beveren, B. Witkamp, L. M. K. Vandersypen, and L. P. Kouwenhoven, Nature (London) \textbf{430}, 431 (2004).

\bibitem{vanRiggelenDoelman2024}
  F. van Riggelen-Doelman, C.-A. Wang, S. L. de Snoo, W. I. L. Lawrie, N. W. Hendrickx, M. Rimbach-Russ, A. Sammak, G. Scappucci, C. D\'eprez, and M. Veldhorst, Nat. Commun. \textbf{15}, 5716 (2024).

\bibitem{David2024Long}
  A. David, A. M. Pazhedath, L. R. Schreiber, T. Calarco, H. Bluhm, and F. Motzoi, \href{https://arxiv.org/abs/2409.07600}{arXiv:2409.07600}.

\end{thebibliography}%


\begin{thebibliography}{15}%
\makeatletter
\providecommand \@ifxundefined [1]{%
 \@ifx{#1\undefined}
}%
\providecommand \@ifnum [1]{%
 \ifnum #1\expandafter \@firstoftwo
 \else \expandafter \@secondoftwo
 \fi
}%
\providecommand \@ifx [1]{%
 \ifx #1\expandafter \@firstoftwo
 \else \expandafter \@secondoftwo
 \fi
}%
\providecommand \natexlab [1]{#1}%
\providecommand \enquote  [1]{``#1''}%
\providecommand \bibnamefont  [1]{#1}%
\providecommand \bibfnamefont [1]{#1}%
\providecommand \citenamefont [1]{#1}%
\providecommand \href@noop [0]{\@secondoftwo}%
\providecommand \href [0]{\begingroup \@sanitize@url \@href}%
\providecommand \@href[1]{\@@startlink{#1}\@@href}%
\providecommand \@@href[1]{\endgroup#1\@@endlink}%
\providecommand \@sanitize@url [0]{\catcode `\\12\catcode `\$12\catcode
  `\&12\catcode `\#12\catcode `\^12\catcode `\_12\catcode `\%12\relax}%
\providecommand \@@startlink[1]{}%
\providecommand \@@endlink[0]{}%
\providecommand \url  [0]{\begingroup\@sanitize@url \@url }%
\providecommand \@url [1]{\endgroup\@href {#1}{\urlprefix }}%
\providecommand \urlprefix  [0]{URL }%
\providecommand \Eprint [0]{\href }%
\providecommand \doibase [0]{http://dx.doi.org/}%
\providecommand \selectlanguage [0]{\@gobble}%
\providecommand \bibinfo  [0]{\@secondoftwo}%
\providecommand \bibfield  [0]{\@secondoftwo}%
\providecommand \translation [1]{[#1]}%
\providecommand \BibitemOpen [0]{}%
\providecommand \bibitemStop [0]{}%
\providecommand \bibitemNoStop [0]{.\EOS\space}%
\providecommand \EOS [0]{\spacefactor3000\relax}%
\providecommand \BibitemShut  [1]{\csname bibitem#1\endcsname}%
\let\auto@bib@innerbib\@empty
\bibitem [{\citenamefont {Magnus}(1954)}]{Magnus1954}%
  \BibitemOpen
  \bibfield  {author} {\bibinfo {author} {\bibfnamefont {W.}~\bibnamefont
  {Magnus}},\ }\href {\doibase 10.1002/cpa.3160070404} {\bibfield  {journal}
  {\bibinfo  {journal} {Communications on Pure and Applied Mathematics}\
  }\textbf {\bibinfo {volume} {7}},\ \bibinfo {pages} {649–673} (\bibinfo
  {year} {1954})}\BibitemShut {NoStop}%
\bibitem [{\citenamefont {Losert}\ \emph {et~al.}(2024)\citenamefont {Losert},
  \citenamefont {Oberl\"ander}, \citenamefont {Teske}, \citenamefont {Volmer},
  \citenamefont {Schreiber}, \citenamefont {Bluhm}, \citenamefont
  {Coppersmith},\ and\ \citenamefont {Friesen}}]{Losert2024Strategies}%
  \BibitemOpen
  \bibfield  {author} {\bibinfo {author} {\bibfnamefont {M.~P.}\ \bibnamefont
  {Losert}}, \bibinfo {author} {\bibfnamefont {M.}~\bibnamefont
  {Oberl\"ander}}, \bibinfo {author} {\bibfnamefont {J.~D.}\ \bibnamefont
  {Teske}}, \bibinfo {author} {\bibfnamefont {M.}~\bibnamefont {Volmer}},
  \bibinfo {author} {\bibfnamefont {L.~R.}\ \bibnamefont {Schreiber}}, \bibinfo
  {author} {\bibfnamefont {H.}~\bibnamefont {Bluhm}}, \bibinfo {author}
  {\bibfnamefont {S.}~\bibnamefont {Coppersmith}}, \ and\ \bibinfo {author}
  {\bibfnamefont {M.}~\bibnamefont {Friesen}},\ }\href {\doibase
  10.1103/PRXQuantum.5.040322} {\bibfield  {journal} {\bibinfo  {journal} {PRX
  Quantum}\ }\textbf {\bibinfo {volume} {5}},\ \bibinfo {pages} {040322}
  (\bibinfo {year} {2024})}\BibitemShut {NoStop}%
\bibitem [{\citenamefont {Goodman}(2020)}]{Goodman2020}%
  \BibitemOpen
  \bibfield  {author} {\bibinfo {author} {\bibfnamefont {J.~W.}\ \bibnamefont
  {Goodman}},\ }\href {\doibase 10.1117/3.2548484} {\emph {\bibinfo {title}
  {Speckle Phenomena in Optics: Theory and Applications, Second Edition}}}\
  (\bibinfo  {publisher} {SPIE},\ \bibinfo {year} {2020})\BibitemShut {NoStop}%
\bibitem [{\citenamefont {Langrock}\ \emph {et~al.}(2023)\citenamefont
  {Langrock}, \citenamefont {Krzywda}, \citenamefont {Focke}, \citenamefont
  {Seidler}, \citenamefont {Schreiber},\ and\ \citenamefont
  {Cywi\ifmmode~\acute{n}\else \'{n}\fi{}ski}}]{Langrock2023Blueprint}%
  \BibitemOpen
  \bibfield  {author} {\bibinfo {author} {\bibfnamefont {V.}~\bibnamefont
  {Langrock}}, \bibinfo {author} {\bibfnamefont {J.~A.}\ \bibnamefont
  {Krzywda}}, \bibinfo {author} {\bibfnamefont {N.}~\bibnamefont {Focke}},
  \bibinfo {author} {\bibfnamefont {I.}~\bibnamefont {Seidler}}, \bibinfo
  {author} {\bibfnamefont {L.~R.}\ \bibnamefont {Schreiber}}, \ and\ \bibinfo
  {author} {\bibfnamefont {L.}~\bibnamefont {Cywi\ifmmode~\acute{n}\else
  \'{n}\fi{}ski}},\ }\href {\doibase 10.1103/PRXQuantum.4.020305} {\bibfield
  {journal} {\bibinfo  {journal} {PRX Quantum}\ }\textbf {\bibinfo {volume}
  {4}},\ \bibinfo {pages} {020305} (\bibinfo {year} {2023})}\BibitemShut
  {NoStop}%
\bibitem [{\citenamefont {Lima}\ and\ \citenamefont
  {Burkard}()}]{Lima2024Superadiabatic}%
  \BibitemOpen
  \bibfield  {author} {\bibinfo {author} {\bibfnamefont {J.~R.~F.}\
  \bibnamefont {Lima}}\ and\ \bibinfo {author} {\bibfnamefont {G.}~\bibnamefont
  {Burkard}},\ }\href {\doibase 10.48550/ARXIV.2408.03173} {}\Eprint
  {http://arxiv.org/abs/arXiv:2408.03173} {arXiv:2408.03173} \BibitemShut
  {NoStop}%
\bibitem [{\citenamefont {Lu}\ \emph {et~al.}(2005)\citenamefont {Lu},
  \citenamefont {Miao},\ and\ \citenamefont {Metcalf}}]{Lu2005}%
  \BibitemOpen
  \bibfield  {author} {\bibinfo {author} {\bibfnamefont {T.}~\bibnamefont
  {Lu}}, \bibinfo {author} {\bibfnamefont {X.}~\bibnamefont {Miao}}, \ and\
  \bibinfo {author} {\bibfnamefont {H.}~\bibnamefont {Metcalf}},\ }\href
  {http://dx.doi.org/10.1103/PhysRevA.71.061405} {\bibfield  {journal}
  {\bibinfo  {journal} {Physical Review A}\ }\textbf {\bibinfo {volume} {71}}
  (\bibinfo {year} {2005})}\BibitemShut {NoStop}%
\bibitem [{\citenamefont {Li}\ \emph {et~al.}(2023)\citenamefont {Li},
  \citenamefont {Spierings},\ and\ \citenamefont {Steinberg}}]{Li2023}%
  \BibitemOpen
  \bibfield  {author} {\bibinfo {author} {\bibfnamefont {K.}~\bibnamefont
  {Li}}, \bibinfo {author} {\bibfnamefont {D.~C.}\ \bibnamefont {Spierings}}, \
  and\ \bibinfo {author} {\bibfnamefont {A.~M.}\ \bibnamefont {Steinberg}},\
  }\href {http://dx.doi.org/10.1103/PhysRevA.108.012615} {\bibfield  {journal}
  {\bibinfo  {journal} {Physical Review A}\ }\textbf {\bibinfo {volume} {108}}
  (\bibinfo {year} {2023})}\BibitemShut {NoStop}%
\bibitem [{\citenamefont {Yang}\ \emph {et~al.}(2013)\citenamefont {Yang},
  \citenamefont {Rossi}, \citenamefont {Ruskov}, \citenamefont {Lai},
  \citenamefont {Mohiyaddin}, \citenamefont {Lee}, \citenamefont {Tahan},
  \citenamefont {Klimeck}, \citenamefont {Morello},\ and\ \citenamefont
  {Dzurak}}]{Yang2013}%
  \BibitemOpen
  \bibfield  {author} {\bibinfo {author} {\bibfnamefont {C.~H.}\ \bibnamefont
  {Yang}}, \bibinfo {author} {\bibfnamefont {A.}~\bibnamefont {Rossi}},
  \bibinfo {author} {\bibfnamefont {R.}~\bibnamefont {Ruskov}}, \bibinfo
  {author} {\bibfnamefont {N.~S.}\ \bibnamefont {Lai}}, \bibinfo {author}
  {\bibfnamefont {F.~A.}\ \bibnamefont {Mohiyaddin}}, \bibinfo {author}
  {\bibfnamefont {S.}~\bibnamefont {Lee}}, \bibinfo {author} {\bibfnamefont
  {C.}~\bibnamefont {Tahan}}, \bibinfo {author} {\bibfnamefont
  {G.}~\bibnamefont {Klimeck}}, \bibinfo {author} {\bibfnamefont
  {A.}~\bibnamefont {Morello}}, \ and\ \bibinfo {author} {\bibfnamefont
  {A.~S.}\ \bibnamefont {Dzurak}},\ }\href
  {http://dx.doi.org/10.1038/ncomms3069} {\bibfield  {journal} {\bibinfo
  {journal} {Nature Communications}\ }\textbf {\bibinfo {volume} {4}} (\bibinfo
  {year} {2013})}\BibitemShut {NoStop}%
\bibitem [{\citenamefont {McKay}\ \emph {et~al.}(2017)\citenamefont {McKay},
  \citenamefont {Wood}, \citenamefont {Sheldon}, \citenamefont {Chow},\ and\
  \citenamefont {Gambetta}}]{McKay2017_virtual}%
  \BibitemOpen
  \bibfield  {author} {\bibinfo {author} {\bibfnamefont {D.~C.}\ \bibnamefont
  {McKay}}, \bibinfo {author} {\bibfnamefont {C.~J.}\ \bibnamefont {Wood}},
  \bibinfo {author} {\bibfnamefont {S.}~\bibnamefont {Sheldon}}, \bibinfo
  {author} {\bibfnamefont {J.~M.}\ \bibnamefont {Chow}}, \ and\ \bibinfo
  {author} {\bibfnamefont {J.~M.}\ \bibnamefont {Gambetta}},\ }\href {\doibase
  10.1103/PhysRevA.96.022330} {\bibfield  {journal} {\bibinfo  {journal} {Phys.
  Rev. A}\ }\textbf {\bibinfo {volume} {96}},\ \bibinfo {pages} {022330}
  (\bibinfo {year} {2017})}\BibitemShut {NoStop}%
\bibitem [{\citenamefont {Nielsen}\ and\ \citenamefont
  {Chuang}(2012)}]{Nielsen2012}%
  \BibitemOpen
  \bibfield  {author} {\bibinfo {author} {\bibfnamefont {M.~A.}\ \bibnamefont
  {Nielsen}}\ and\ \bibinfo {author} {\bibfnamefont {I.~L.}\ \bibnamefont
  {Chuang}},\ }\href {\doibase 10.1017/cbo9780511976667} {\emph {\bibinfo
  {title} {Quantum Computation and Quantum Information: 10th Anniversary
  Edition}}}\ (\bibinfo  {publisher} {Cambridge University Press},\ \bibinfo
  {year} {2012})\BibitemShut {NoStop}%
\bibitem [{\citenamefont {Swinbank}\ and\ \citenamefont
  {James~Purser}(2006)}]{Swinbank2006}%
  \BibitemOpen
  \bibfield  {author} {\bibinfo {author} {\bibfnamefont {R.}~\bibnamefont
  {Swinbank}}\ and\ \bibinfo {author} {\bibfnamefont {R.}~\bibnamefont
  {James~Purser}},\ }\href {\doibase 10.1256/qj.05.227} {\bibfield  {journal}
  {\bibinfo  {journal} {Quarterly Journal of the Royal Meteorological Society}\
  }\textbf {\bibinfo {volume} {132}},\ \bibinfo {pages} {1769–1793} (\bibinfo
  {year} {2006})}\BibitemShut {NoStop}%
\bibitem [{\citenamefont {Gonzalez}(2009)}]{Gonzlez2009}%
  \BibitemOpen
  \bibfield  {author} {\bibinfo {author} {\bibfnamefont {A.}~\bibnamefont
  {Gonzalez}},\ }\href {\doibase 10.1007/s11004-009-9257-x} {\bibfield
  {journal} {\bibinfo  {journal} {Mathematical Geosciences}\ }\textbf {\bibinfo
  {volume} {42}},\ \bibinfo {pages} {49–64} (\bibinfo {year}
  {2009})}\BibitemShut {NoStop}%
\bibitem [{\citenamefont {Mao}\ \emph {et~al.}(2022)\citenamefont {Mao},
  \citenamefont {Liu}, \citenamefont {Zeb}, \citenamefont {Lu}, \citenamefont
  {Liu}, \citenamefont {Poole}, \citenamefont {Song},\ and\ \citenamefont
  {Barrios}}]{Mao2022}%
  \BibitemOpen
  \bibfield  {author} {\bibinfo {author} {\bibfnamefont {Y.}~\bibnamefont
  {Mao}}, \bibinfo {author} {\bibfnamefont {G.}~\bibnamefont {Liu}}, \bibinfo
  {author} {\bibfnamefont {K.}~\bibnamefont {Zeb}}, \bibinfo {author}
  {\bibfnamefont {Z.}~\bibnamefont {Lu}}, \bibinfo {author} {\bibfnamefont
  {J.}~\bibnamefont {Liu}}, \bibinfo {author} {\bibfnamefont {P.~J.}\
  \bibnamefont {Poole}}, \bibinfo {author} {\bibfnamefont {C.-Y.}\ \bibnamefont
  {Song}}, \ and\ \bibinfo {author} {\bibfnamefont {P.}~\bibnamefont
  {Barrios}},\ }\href {\doibase 10.3390/photonics9100695} {\bibfield  {journal}
  {\bibinfo  {journal} {Photonics}\ }\textbf {\bibinfo {volume} {9}} (\bibinfo
  {year} {2022}),\ 10.3390/photonics9100695}\BibitemShut {NoStop}%
\bibitem [{\citenamefont {Wang}\ \emph {et~al.}(2024)\citenamefont {Wang},
  \citenamefont {John}, \citenamefont {Tidjani}, \citenamefont {Yu},
  \citenamefont {Ivlev}, \citenamefont {Déprez}, \citenamefont {van
  Riggelen-Doelman}, \citenamefont {Woods}, \citenamefont {Hendrickx},
  \citenamefont {Lawrie}, \citenamefont {Stehouwer}, \citenamefont
  {Oosterhout}, \citenamefont {Sammak}, \citenamefont {Friesen}, \citenamefont
  {Scappucci}, \citenamefont {de~Snoo}, \citenamefont {Rimbach-Russ},
  \citenamefont {Borsoi},\ and\ \citenamefont {Veldhorst}}]{Wang2024}%
  \BibitemOpen
  \bibfield  {author} {\bibinfo {author} {\bibfnamefont {C.-A.}\ \bibnamefont
  {Wang}}, \bibinfo {author} {\bibfnamefont {V.}~\bibnamefont {John}}, \bibinfo
  {author} {\bibfnamefont {H.}~\bibnamefont {Tidjani}}, \bibinfo {author}
  {\bibfnamefont {C.~X.}\ \bibnamefont {Yu}}, \bibinfo {author} {\bibfnamefont
  {A.~S.}\ \bibnamefont {Ivlev}}, \bibinfo {author} {\bibfnamefont
  {C.}~\bibnamefont {Déprez}}, \bibinfo {author} {\bibfnamefont
  {F.}~\bibnamefont {van Riggelen-Doelman}}, \bibinfo {author} {\bibfnamefont
  {B.~D.}\ \bibnamefont {Woods}}, \bibinfo {author} {\bibfnamefont {N.~W.}\
  \bibnamefont {Hendrickx}}, \bibinfo {author} {\bibfnamefont {W.~I.~L.}\
  \bibnamefont {Lawrie}}, \bibinfo {author} {\bibfnamefont {L.~E.~A.}\
  \bibnamefont {Stehouwer}}, \bibinfo {author} {\bibfnamefont {S.~D.}\
  \bibnamefont {Oosterhout}}, \bibinfo {author} {\bibfnamefont
  {A.}~\bibnamefont {Sammak}}, \bibinfo {author} {\bibfnamefont
  {M.}~\bibnamefont {Friesen}}, \bibinfo {author} {\bibfnamefont
  {G.}~\bibnamefont {Scappucci}}, \bibinfo {author} {\bibfnamefont {S.~L.}\
  \bibnamefont {de~Snoo}}, \bibinfo {author} {\bibfnamefont {M.}~\bibnamefont
  {Rimbach-Russ}}, \bibinfo {author} {\bibfnamefont {F.}~\bibnamefont
  {Borsoi}}, \ and\ \bibinfo {author} {\bibfnamefont {M.}~\bibnamefont
  {Veldhorst}},\ }\href {\doibase 10.1126/science.ado5915} {\bibfield
  {journal} {\bibinfo  {journal} {Science}\ }\textbf {\bibinfo {volume}
  {385}},\ \bibinfo {pages} {447–452} (\bibinfo {year} {2024})}\BibitemShut
  {NoStop}%
\bibitem [{\citenamefont {Yoneda}\ \emph {et~al.}(2017)\citenamefont {Yoneda},
  \citenamefont {Takeda}, \citenamefont {Otsuka}, \citenamefont {Nakajima},
  \citenamefont {Delbecq}, \citenamefont {Allison}, \citenamefont {Honda},
  \citenamefont {Kodera}, \citenamefont {Oda}, \citenamefont {Hoshi},
  \citenamefont {Usami}, \citenamefont {Itoh},\ and\ \citenamefont
  {Tarucha}}]{Yoneda2017}%
  \BibitemOpen
  \bibfield  {author} {\bibinfo {author} {\bibfnamefont {J.}~\bibnamefont
  {Yoneda}}, \bibinfo {author} {\bibfnamefont {K.}~\bibnamefont {Takeda}},
  \bibinfo {author} {\bibfnamefont {T.}~\bibnamefont {Otsuka}}, \bibinfo
  {author} {\bibfnamefont {T.}~\bibnamefont {Nakajima}}, \bibinfo {author}
  {\bibfnamefont {M.~R.}\ \bibnamefont {Delbecq}}, \bibinfo {author}
  {\bibfnamefont {G.}~\bibnamefont {Allison}}, \bibinfo {author} {\bibfnamefont
  {T.}~\bibnamefont {Honda}}, \bibinfo {author} {\bibfnamefont
  {T.}~\bibnamefont {Kodera}}, \bibinfo {author} {\bibfnamefont
  {S.}~\bibnamefont {Oda}}, \bibinfo {author} {\bibfnamefont {Y.}~\bibnamefont
  {Hoshi}}, \bibinfo {author} {\bibfnamefont {N.}~\bibnamefont {Usami}},
  \bibinfo {author} {\bibfnamefont {K.~M.}\ \bibnamefont {Itoh}}, \ and\
  \bibinfo {author} {\bibfnamefont {S.}~\bibnamefont {Tarucha}},\ }\href
  {\doibase 10.1038/s41565-017-0014-x} {\bibfield  {journal} {\bibinfo
  {journal} {Nature Nanotechnology}\ }\textbf {\bibinfo {volume} {13}},\
  \bibinfo {pages} {102–106} (\bibinfo {year} {2017})}\BibitemShut {NoStop}%
\end{thebibliography}%
\end{document}